\documentclass[12pt]{iopart}
\pdfoutput=1
\usepackage{iopams}
\usepackage{epsfig}   
\usepackage{graphics}
\usepackage[numbers,square,sort]{natbib}

\newcommand{\om}{\Omega_{m}}
\newcommand{\ok}{\Omega_{k}}
\newcommand{\ola}{\Omega_{\Lambda}}
\newcommand{\laa}{l_{\rm a}}

\begin{document}

\title%[Has the universal expansion gone from deceleration to acceleration?]
%{Is the universal expansion accelerating now and was it decelerating earlier?}
%{Has the universal expansion gone from deceleration to acceleration?}
{Model independent constraints on the cosmological expansion rate}

\author{Edvard M\"ortsell$^1$ and Chris Clarkson$^2$}

\address{$^1$ Department of Physics, Stockholm University, AlbaNova
         University Center \\ S--106 91 Stockholm, Sweden}
\address{$^2$ Cosmology and Gravity Group, Department of Mathematics and 
Applied Mathematics, University of Cape Town, Rondebosch 7701, Cape Town, South Africa}
         
\ead{\mailto{edvard@physto.se}, \mailto{chris.clarkson@uct.ac.za}}

\begin{abstract}
We investigate what current cosmological data tells us about the
cosmological expansion rate in a model independent way. Specifically,
we study if the expansion was decelerating at high redshifts and is
accelerating now, without referring to any model for the energy
content of the universe, nor to any specific theory of gravity. This
differs from most studies of the expansion rate which, e.g., assumes
some underlying parameterised model for the dark energy component of
the universe. To accomplish this, we have devised a new method to probe
the expansion rate without relying on such assumptions.

Using only supernova data, we conclude that there is little doubt that
the universe has been accelerating at late times. However, contrary to
some previous claims, we can not determine if the universe was
previously decelerating. For a variety of methods used for
constraining the expansion history of the universe, acceleration is
detected from supernovae alone at $>5\sigma$, regardless of the
curvature of the universe. Specifically, using a Taylor expansion of
the scale factor, acceleration today is detected at $>12\sigma$. If we
also include the ratio of the scale of the baryon acoustic
oscillations as imprinted in the cosmic microwave background and in
the large scale distribution of galaxies, it is evident from the data
that the expansion decelerated at high redshifts, but only with the
assumption of a flat or negatively curved universe.
\end{abstract}

\noindent{\it Keywords}: dark energy theory, supernova type Ia
%\maketitle

%%%%%%%%%%%%%%%%%%%%%%%%%%%%%%%%%%%%%%%%%%%%%%%%%%%%%%%%%%%%%%%%%%%%%%%
%%%%%%%%%%%%%%%%%%%%%%%%%%%%%%%%%%%%%%%%%%%%%%%%%%%%%%%%%%%%%%%%%%%%%%%
\section{Introduction}
It is becoming generally acknowledged that the observed
redshift-distance relation from Type Ia supernovae (SNe Ia), together
with the scales of baryon acoustic oscillations as observed in the
distribution of galaxies on large scales (BAO) and the temperature
anisotropies in the cosmic microwave background (CMB), implies that
the current energy density in the universe is dominated by dark
energy, here defined as a component with an equation of state,
$w=p/\rho<-1/3$. Current cosmological data are consistent with the
standard -- or concordance -- cosmological model, where $\om=0.3$ and
$\ola=0.7$, i.e., the dark energy is explained in terms of a
cosmological constant or vacuum energy with $w=-1$
\citep[e.g.,][]{2008arXiv0803.0547K,2007ApJ...666..716D,
2007ApJ...659...98R,2006A&A...447...31A,2005ApJ...633..560E,
2005MNRAS.362..505C}. For the concordance model, the universal
expansion is accelerating at redshifts lower than $z\sim 0.7$ and
decelerating at higher redshifts. 

Most analyses which attempt to infer something about the expansion
history of the universe rely on a specific model, e.g., a dark energy
`fluid' or a modification of general relativity, which has one or two
parameters of interest~\citep{2006IJMPD..15.1753C}. Consequently, much
of our knowledge of the expansion history of the universe has these
parameterisations hard-wired into our conclusions, which may leave a
large space of possible expansion histories unexplored. Alongside
these analyses, it is therefore constructive to try to assert model
independent statements where we can. The question we address in this
paper is to what extent we can infer changes in the expansion rate
without referring to any theory of gravity or model for the energy
content of the universe. That is; what is the history of the universal
expansion?

Unfortunately, it is difficult to give a definite answer to this
question since two important assumptions are implicit in all
discussion of this kind. It is known that spherically symmetric
void~-- or Hubble bubble~-- models may explain the anomalous Hubble
diagram whilst always maintaining a decelerating expansion rate, at
the price of violating the Copernican
principle~\citep{2008JCAP...09..016G,2008PhRvD..78h3511V,2008PhRvL.100s1302C,
2006yosc.conf....9B,2008GReGr..40..269S,2008JCAP...04..003G,2008arXiv0807.1443C,
2008arXiv0807.2891B,2008PhRvD..78d3504Z,2008arXiv0810.4939G}. Ideally
we should test the Copernican principle in a model independent
way~\citep{2008PhRvL.101a1301C, 2008PhRvL.100s1303U}, and so rule
these models out. A further assumption in the standard model~--
insofar as determining the expansion dynamics is concerned~-- is that
that the universe is smooth enough at small distances to be described
by a perfectly homogeneous and isotropic
model~\citep{2008arXiv0810.4484S,2007PhRvD..76h3011L,
2008arXiv0801.3420L,2008GReGr..40..467B,2008JCAP...01..013B,2008JCAP...04..026R,
2008arXiv0808.1161L,2008arXiv0809.2107R,2007arXiv0704.1734C,2008arXiv0812.2872R}. At
best this gives a small error to all our considerations; at worst,
many of our conclusions might be wrong.

Since it was traditionally thought that the expansion rate would be
decelerating, we measure acceleration with the deceleration parameter,
$q$, defined by
\begin{equation}
  q \equiv -\frac{\ddot a a}{\dot a^2}=-\frac{\ddot a}{a H^2}\, ,
\end{equation}
where $a$ is the scale factor, dots denote derivatives with respect to
time and the Hubble parameter, $H$, is defined as $H\equiv \dot
a/a$. Because of the sign convention, negative values for $q$
correspond to acceleration.  Traditionally, the goal when observing
the expansion of the universe was to constrain two parameters, the
current values of the Hubble parameter -- or the Hubble constant --
$H_0$, and the deceleration parameter, $q_0$.  The first signs of an
accelerated expansion came 10 years ago with the observations that
distant SN Ia appear dimmer than expected in a universe with constant
or decelerated expansion velocity
\citep{1998AJ....116.1009R,1998ApJ...507...46S,1999ApJ...517..565P}. 
However, the acceleration was only evaluated in terms of a model with
$\om$ and $\ola$, in which $q_0 =
\om/2-\ola$, see Eq.~(\ref{eq:qzok}).

In 2002, \citet{2002ApJ...569...18T} studied the change in the
expansion velocity -- without referring to the energy content or
theory of gravity -- by using a step model for the deceleration
parameter where $q$ had one constant value at low and intermediate
redshifts and another constant value at high redshifts. They
demonstrated, that the SN Ia data at the time showed a strong
preference for acceleration today and deceleration in the past, if the
transition redshift was set (by hand) to $z=0.4-0.6$. However, this
conclusion relied heavily on the observed magnitude of a single
supernova, SN1997ff, the most distant SN Ia observed at $z=1.755$.
Being very distant, it is especially susceptible to systematic effects
that may reduce its cosmological utility.  One such effect is
gravitational lensing that has been shown to brighten SN1997ff by
$\sim 0.15$ magnitudes \citep{2006ApJ...639..991J}.

Also, in \citet{2006ApJ...649..563S}, it was shown that marginalising
over the transition redshift, considerably relaxed the constraints on
the expansion history. Specifically, using the so called ``gold''
dataset consisting of 157 SNe Ia \citep{2004ApJ...607..665R}, the
authors only found strong evidence for acceleration at some epoch,
not necessarily at $z<0.1$, and that $q$ was higher in the
past.

In 2004, \citet{2004ApJ...607..665R}, the gold dataset was used to
constrain a Taylor expansion of $q(z)$,
\begin{equation}
  q(z)=q_0 + zq'(z=0)\, ,
\end{equation}
where the prime denotes a derivative with respect to redshift. The
result found was that $q_0\lesssim -0.3$ and $q'\gtrsim 0$ at $95\,\%$
confidence level (CL). The evidence that $q'\gtrsim 0$, was
interpreted as evidence for deceleration at higher redshifts. However,
the Taylor expansion should only be meaningful for $z<1$, up to which
the parameter space $[q_0,q']$ still allows for acceleration at
$95\,\%$ CL. Also, the fit is done using data at $z>1$, for which
higher order terms in the Taylor expansion should be important.

In, \citet{2006JCAP...09..002E}, the same data was found to be
consistent with a constant negative deceleration parameter, and in
\citet{2007MNRAS.375.1510R}, the SN data was combined with X-ray
cluster gas mass fraction measurements to constrain the deceleration
parameter, $q$, as well as the next order derivative of the scale
factor as decoded in the jerk parameter, $j$. See also
\citet{2008ApJ...677....1D} for an alternative approach to 
constraining the acceleration history of the universe.

In this paper, we follow the spirit of previous work, in that we
strive to make as few assumptions as possible regarding the theory of
gravity or the energy content of the universe when inferring the state
of the expansion velocity of the universe. In fact, the only
assumptions used in this paper, other than those mentioned, is that
SNe Ia are standardisable candles and that the observed
inhomogeneities in the large scale distribution of galaxies and the
anisotropies in the temperature of the CMB reflects the same physical
scale. One feature we discuss in particular is the role of curvature
in determining the acceleration, as there are significant degeneracies
between curvature and acceleration, in a similar vein which exists
between curvature and the dark energy equation of state
$w$~\cite{2007JCAP...08..011C,2008GReGr..40..285H}.

In Sec.~\ref{sec:acc}, we discuss acceleration and deceleration within
the standard model, as well as possible observational measures of the
deceleration parameter. In Sec.~\ref{sec:data}, we present the two
sources of data used in this paper, and in Sec.~\ref{sec:method}, we
present a new method for inferring the expansion history of the
universe, together with our results for $q(z)$. Our results are
summarised in Sec.~\ref{sec:summary}. In short, we conclude that the
evidence for late time acceleration of the universal expansion is very
strong, regardless of the method used to measure $q(z)$, whereas
deceleration at high redshifts can still be avoided given current
data.

%%%%%%%%%%%%%%%%%%%%%%%%%%%%%%%%%%%%%%%%%%%%%%%%%%%%%%%%%%%%%%%%%%%%%%%
%%%%%%%%%%%%%%%%%%%%%%%%%%%%%%%%%%%%%%%%%%%%%%%%%%%%%%%%%%%%%%%%%%%%%%%
\section{Acceleration and deceleration}\label{sec:acc}
\begin{figure}
\begin{center}
\includegraphics[angle=0,width=.65\textwidth]{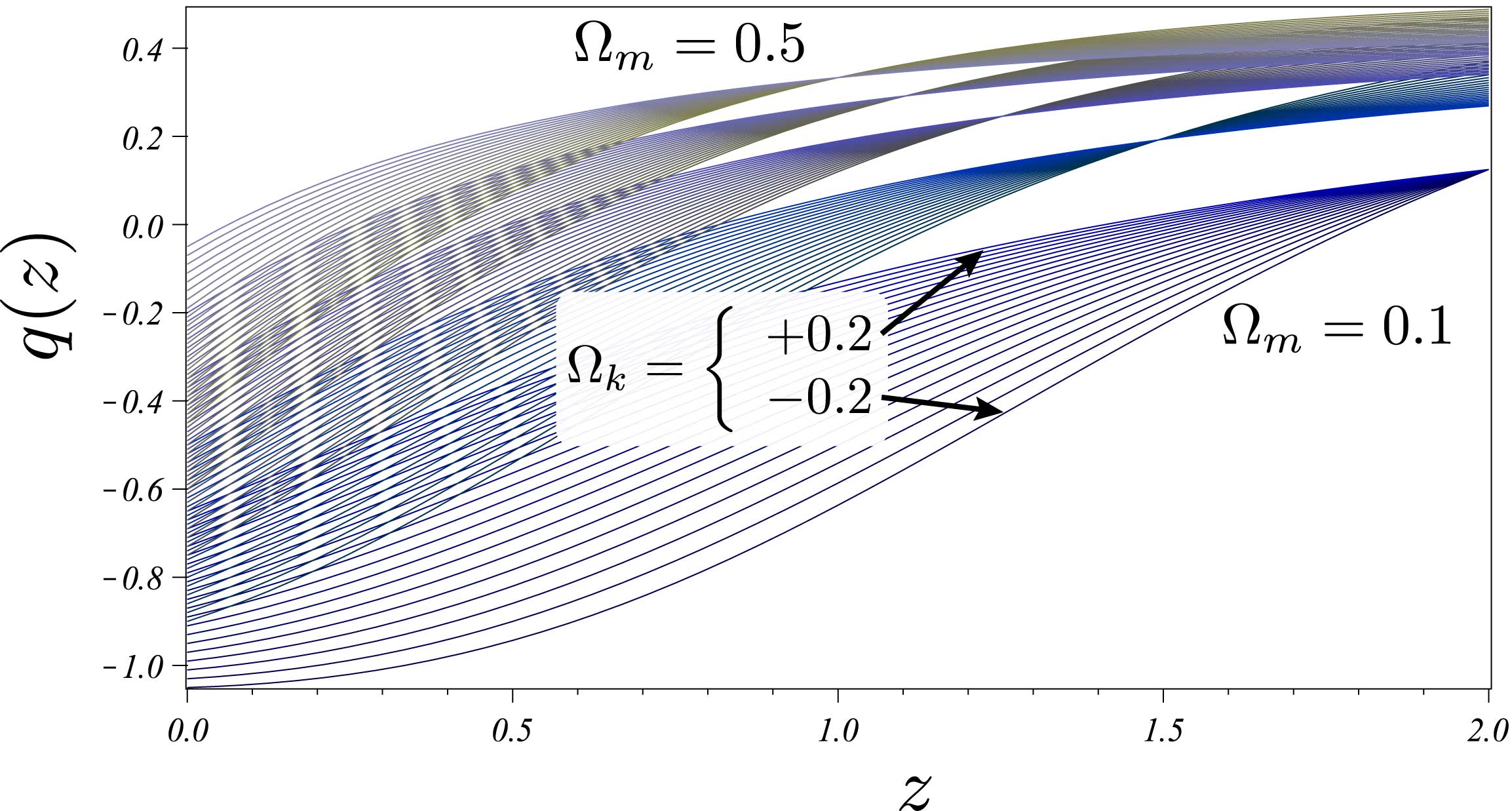}
\caption{\label{fig:q1} The range of behaviour available to $q(z)$ 
for the $\Lambda$CDM model.}
\end{center}
\end{figure}

%%%%%%%%%%%%%%%%%%%%%%%%%%%%%%%%%%%%%%%%%%%%%%%%%%%%%%%%%%%%%%%%%%%%%%%
First, let us calculate what to expect for $q(z)$ in the standard
$\Lambda$CDM model, with matter density, $\om$, and a cosmological
constant, $\ola = 1-\om-\ok$. The acceleration equation is given by
\begin{equation}
  \frac{\ddot a}{a}=-\frac{4\pi G}{3}(\epsilon + 3p)=
  -\frac{4\pi G}{3}(\rho_m -2\rho_\Lambda)\, .
\end{equation}
This gives 
\begin{equation}\label{eq:qzok}
  %q(z) = \frac{\om(1+z)^3/2-1+\om}{\om(1+z)^3+1-\om} \, .
  q (z) =\frac{1}{2}\frac {\Omega_{{m}} \left( 1
+z \right) ^{3}-2(1-\Omega_{{m}}-\Omega_{{k}})}{\Omega_{{m}} \left( 1+z \right) ^{3
}+\Omega_{{k}} \left( 1+z \right) ^{2}+1-\Omega_{{m}}-\Omega_{{k}}}\, .
\end{equation}
In the infinite future ($z=-1$), we have $q=-1$, and as
$z\to\infty$\footnote{In fact, this approximation is only valid after
the universe became matter dominated at $z\sim 3000$.}, $q=0.5$.  The
transition redshift between acceleration and deceleration is given by
$z_t = [2(1-\om)/\om]^{1/3}-1$ when $\ok=0$. For the flat concordance
cosmology with $\om = 0.3$, we have $z_t\sim 0.7$. One of the main
problems we have in determining acceleration as a function of
redshift, is the wide range of behaviours $q(z)$ can exhibit,
especially if curvature is present. In Fig~\ref{fig:q1}, we show the
range of values $q(z)$ can take on within the $\Lambda$CDM paradigm by
varying the constants $\Omega_m$ and $\Omega_k$.
%----------------------------------------------------------------------
%----------------------------------------------------------------------
With more exotic dark energy models the variation can be much more
elaborate and $q(z)$ can be regarded as a free function to be
constrained by observations. Then, once $q(z)$ is determined, the dark
energy equation of state, $w(z)$, can in principle be determined from
a first-order differential equation. However, since the integration
constant (i.e., $H_0$), as well as $\ok$ and $\om$ are arbitrary,
there is a three parameter family of dark energy models which can give
rise to the same $q(z)$.

Normalising the scale factor to be unity today, $a = (1+z)^{-1}$, we
may derive $\dot z = -(1+z)H$, and
\begin{equation}\label{eq:q}
	q(z) = -\frac{\dot H}{H^2}-1 =
	\frac{H'}{H}(1+z)-1=(1+z)\left[\ln\frac{H}{1+z}\right]'\, .
\end{equation}
In principle, Eq.~(\ref{eq:q}) can be used to obtain the value of the
deceleration parameter as a function of redshift.  Unfortunately, it
is very difficult to measure the Hubble parameter, $H(z)$, let alone
its derivative. As an example, from Fig.~\ref{fig:magdiff}, it can be
understood why it is difficult to differentiate noisy SN Ia data in
order to obtain $H(z)$. Note however that with future BAO data, it may
be possible to measure $H(z)$ directly \citep{2003ApJ...598..720S}.

Rearranging Eq.~(\ref{eq:q}), we can write
\begin{equation}
	\frac{H'}{H} = \frac{1+q(z)}{1+z} \rightarrow
	\int_{z_1}^{z_2}{\frac{d(\ln
	H)}{dz}dz}=\int_{z_1}^{z_2}{\frac{1+q(z)}{1+z}dz}\, .
\end{equation}
If the universe is accelerating between redshifts $z_1$ and $z_2$, we
have
\begin{equation}
	\ln\left(\frac{H_2}{H_1}\right) <
	\ln\left(\frac{1+z_2}{1+z_1}\right) \rightarrow
	\frac{(1+z_2)}{H_2}>\frac{(1+z_1)}{H_1}\, .
\end{equation}
When we have acceleration, $H(z)$ thus grows with redshift faster than
$(1+z)$. This is easy to understand since if $\dot a=H/(1+z)$ is
increasing with redshift, $\dot a$ is decreasing with time and $\ddot
a <0$, corresponding to acceleration. For deceleration, $H(z)$ grows
slower than $(1+z)$. For a matter dominated universe, $H =
H_0\sqrt{\om (1+z)^3}\propto (1+z)^{1.5}$, i.e., the expansion is
decelerating. For an empty universe, $H = H_0\sqrt{\ok (1+z)^2}\propto
(1+z)$, i.e., the expansion velocity is constant.  Note however that
an empty universe is not the only possible solution for a constant
expansion velocity -- any universe where the total energy density
scales as $(1+z)^2$ also gives a constant expansion.

The comoving coordinate distance is given by
\begin{equation}
  d_c(z) = \int_{0}^{z}{\frac{dz}{H(z)}}=
  \frac{1}{H_0}\int_{0}^{z}{\exp{\left[-\int_0^{v}\frac{[1+q(u)]du}{(1+u)}\right]}dv}
  \, ,
\end{equation}
and the luminosity distance, which is the relevant quantity for SN Ia
observations, is given by
\begin{equation}
  d_L(z) = \frac{1+z}{H_0\sqrt{-\ok}}\sin{\left[ \sqrt{-\ok} H_0 d_c(z)\right]} \, .
\end{equation}
The angular diameter distance, relevant for the scale of BAO and CMB,
is given by $d_A = d_L/(1+z)^2$.  Although having the same constant
expansion rate, we would therefore measure different luminosity and
angular diameter distances in the flat and open non-accelerating
cases. Determining acceleration at a given redshift is then the same
as determining if the function $ [(1+z)D'/H_0\sqrt{1-|\Omega_k|
D^2}]$, where $D(z)=d_L(z)H_0/(1+z)$, is an increasing function of
redshift.

%%%%%%%%%%%%%%%%%%%%%%%%%%%%%%%%%%%%%%%%%%%%%%%%%%%%%%%%%%%%%%%%%%%%%%%
%%%%%%%%%%%%%%%%%%%%%%%%%%%%%%%%%%%%%%%%%%%%%%%%%%%%%%%%%%%%%%%%%%%%%%%
\section{Data}\label{sec:data}
%%%%%%%%%%%%%%%%%%%%%%%%%%%%%%%%%%%%%%%%%%%%%%%%%%%%%%%%%%%%%%%%%%%%%%%
In the last decade, there has been a formidable progress in using
cosmological data to constrain the expansion history and the energy
content of the universe. Observations include, but are not restricted
to, SNe Ia, BAO, CMB, weak gravitational lensing and galaxy cluster
number counts. In this paper, we make use of SN Ia data together with
a combination of CMB and BAO observations that only depends on the
expansion history of the universe, not the energy content.

%%%%%%%%%%%%%%%%%%%%%%%%%%%%%%%%%%%%%%%%%%%%%%%%%%%%%%%%%%%%%%%%%%%%%%%
\subsection{Type Ia supernova data}
The Union08 data set \cite{2008arXiv0804.4142K} is a compilation of
SNe Ia from, e.g., the Supernova Legacy Survey, ESSENCE survey and
HST. After selection cuts, the data set amounts to 307 SNe Ia,
spanning a redshift range of $0\lesssim z \lesssim 1.55$, analysed in
a homogenous fashion using the spectral-template-based fit method
SALT.

%%%%%%%%%%%%%%%%%%%%%%%%%%%%%%%%%%%%%%%%%%%%%%%%%%%%%%%%%%%%%%%%%%%%%%%
\subsection{Baryon acoustic oscillations and the cosmic microvawe background}
The distances measured to the CMB decoupling epoch at $z_*\sim 1090$
and to the BAO at $z=[0.2, 0.35]$ depend on the physical scale of the
acoustic oscillations at decoupling, and thus on the matter and baryon
density. Specifically, the position of the first peak in the CMB power
spectrum, which represents the angular scale of the sound horizon at
decoupling, is given by,
\begin{equation} 
  \laa \approx \pi \frac{d_A(z_{*}) (1+z_{*})}{r_{\rm s}(z_{*})}
\end{equation}
where the comoving
sound horizon at recombination,
\begin{equation} 
  r_{\rm s}(z_*) = \int_{z_{*}}^\infty \frac{c_s(z)}{H(z)}dz\, ,
\end{equation}
depends on the speed of sound, $c_s$, in the early universe. The
observed scale of the BAO is given by $r_{\rm s}(z_{*})/D_V$,
where the so called dilation scale, $D_V$, is combined from angular
diameter and radial distances according to
\begin{equation} 
  D_V(z) = \left[(1+z)^2d_A^2 \frac{cz}{H(z)}\right]^{1/3} \, .
\end{equation}

Since we want to infer the expansion history with minimal assumptions
regarding the energy density, we take a conservative approach and use
the ratio of the observed scales in the CMB and the BAO. This ratio
does not depend on the physical size of the sound horizon at
decoupling but only on the assumption that the BAO and CMB reflects
the {\em same} physical size. \citet{2007MNRAS.381.1053P} derives
\begin{eqnarray}
  \frac{d_A(z_{*}) (1+z_{*})}{D_V(z=0.2)} &=& 19.04\pm 0.58\nonumber\\
  \frac{d_A(z_{*}) (1+z_{*})}{D_V(z=0.35)} &=& 10.52\pm 0.32\label{eq:BAO} \, .
\end{eqnarray}
The measurements (from the 2dFGRS and SDSS, respectively, combined
with 3 year WMAP data) are correlated with correlation coefficient
$\rho = 0.39$. Using WMAP 5 year data instead of 3 year data gives
close to identical results, when combined with the BAO data.

The constraints given by Eq.~(\ref{eq:BAO}), individually place very
weak constraints on acceleration. Also, these constraints are highly
degenerate with the curvature.
%----------------------------------------------------------------------
\begin{figure}
\begin{center}
\includegraphics[angle=0,width=.75\textwidth]{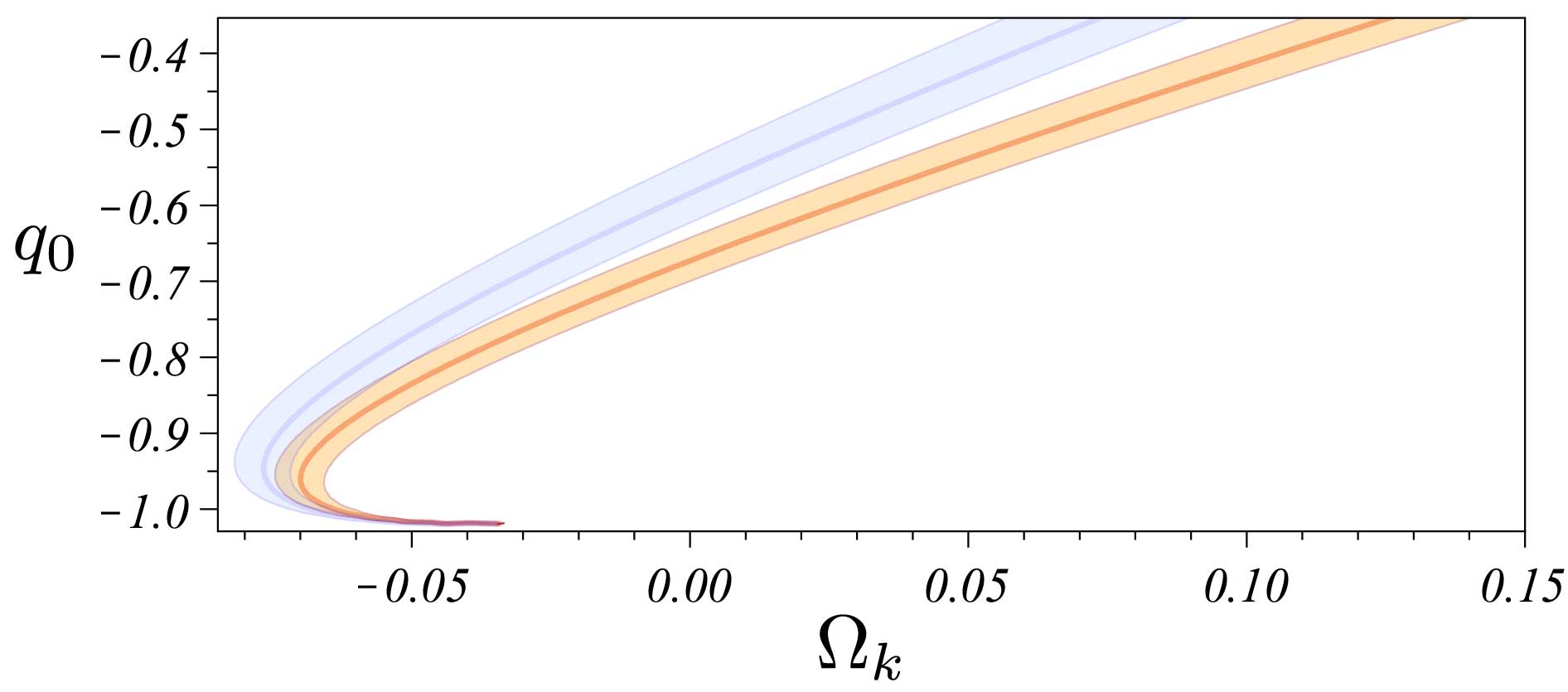}
\caption{\label{fig:BAO} Best fit values of ${d_A(z_{*})}/{D_V(z)}$ at $z=0.2$ 
(red) and $z=0.35$ (blue), together with their $1\sigma$ errors, as
measured from the scale imprinted in the CMB and BAO, shown in the
parameter space for $\Lambda$CDM.}
\end{center}
\end{figure}
%----------------------------------------------------------------------
In Fig.~\ref{fig:BAO}, we show the CMB/BAO constraints in the
$\ok-q_0$ parameter space for the $\Lambda$CDM model, with their
$1\sigma$ errors. It is clear that they require a large, negative,
deceleration parameter to be consistent with each other at $1\sigma$
CL, irrespective of the curvature. Note also that a flat model is
ruled out at this level.  At $2\sigma$ CL however, individual
constraints on $q_0$ and $\Omega_k$ becomes very weak, and the
combined constraints are limited by the strong degeneracy exhibited
between $\Omega_k$ and $q_0$. Nevertheless, the data unambiguosly
shows that $\Omega_k>-0.1$, i.e., the universe is not strongly
overclosed.

%%%%%%%%%%%%%%%%%%%%%%%%%%%%%%%%%%%%%%%%%%%%%%%%%%%%%%%%%%%%%%%%%%%%%%%
%%%%%%%%%%%%%%%%%%%%%%%%%%%%%%%%%%%%%%%%%%%%%%%%%%%%%%%%%%%%%%%%%%%%%%%
\section{Methods and results}\label{sec:method}
%%%%%%%%%%%%%%%%%%%%%%%%%%%%%%%%%%%%%%%%%%%%%%%%%%%%%%%%%%%%%%%%%%%%%%%
In this section, we discuss how to determine the change of the
expansion rate, or $q(z)$, without assuming a model for the energy
content of the universe or a specific gravitational theory. We are
thus limited to using geometrical data as opposed to methods that are
sensitive to the growth of structure in the universe.

First, we discuss some difficulties that occur when trying to find
$q(z)$ by comparing observed magnitudes of SNe Ia directly. This
method has the advantage that it can, in principle, detect a change in
the expansion rate without somewhat ad-hoc parameterisations of
$q(z)$. Some of these difficulties have been realised by
\citet{2008arXiv0810.4484S} -- in particular the reliance on low
redshift SNe Ia.

Alternatively we may parametrise the expansion with piecewise,
constant accelerations, i.e., we assume that $q(z)$ can vary between
redshift bins, but is constant within the bins. This is the approach
introduced by \citet{2002ApJ...569...18T}. We investigate this further
below, and find that it suffers from strong degeneracies with
curvature when including CMB data. We also investigate alternative
parameterisations of $q(z)$, displaying the same
degeneracies. However, we show that SN Ia data alone, show that
$q(z=0)<0$, irrespective of curvature.

We then present a method for determining acceleration, first employed
in \citet{2007ApJ...659...98R}, that relies on the fact that $\dot
a^{-1}=(1+z)/H(z)$ is an increasing function when the expansion is
accelerating which seems to the case at low redshifts. However, since
the method relies on differentiating noisy SN Ia data, results at high
redshifts have very large uncertainties. Finally we present a new
method which relies on a Taylor expansion around an arbitrary
`sliding' redshift.

\subsection{Inferring acceleration from $m(z)$}
In SN Ia cosmology, the validity of a given cosmological model is
tested by comparing observed peak SN Ia magnitudes with the
theoretical magnitudes for the given model,
\begin{equation}
  m(z) = M + 5\log_{\rm 10}\left[\frac{d_L(z)}{1\,{\rm Mpc}}\right] + 25\, ,
\end{equation}
where $M$ is the absolute SN Ia magnitude. The normalisation of the
magnitude (containing, e.g., $M$ and $H_0$) is usually marginalised
over, and constraints are derived by examining the redshift evolution
of the magnitudes.

Often, one presents SN Ia data in the form of the difference between
the observed peak magnitudes, and the theoretical magnitudes in an
empty universe, $m_e$, where the difference is normalised to be zero
at low redshifts. Since an empty universe is neither accelerating nor
decelerating, it is sometimes claimed that one can infer the state of
the universal expansion from a quick visual inspection of this
difference. One common claim (at talks and discussions, if not in
papers) is that negative values of this difference, $m-m_e<0$, shows
that the universe is decelerating and positive values that it is
accelerating. Another inconsistent, but nevertheless common claim, is
that if the difference increases with redshift, the universal
expansion is accelerating. If the difference is decreasing, the
expansion is decelerating. That is, the claim is that we can infer the
state of the expansion velocity by studying the sign of the derivative
of the difference with respect to redshift. Let us investigate these
claims.

\begin{figure}
\begin{center}
\includegraphics[angle=0,width=.65\textwidth]{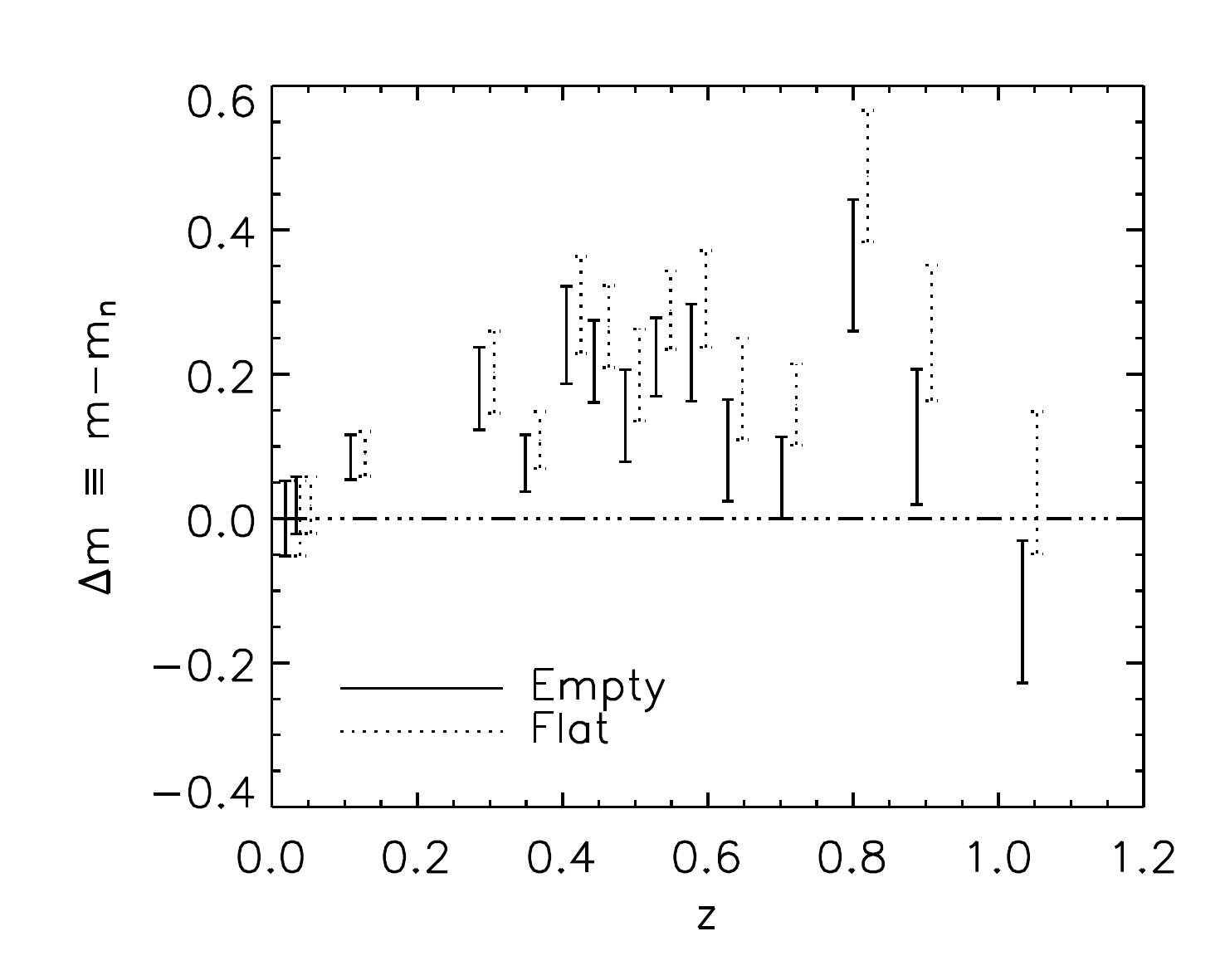}
\caption{\label{fig:magdiff} The differences between observed 
SN Ia magnitudes and the expected magnitudes for empty and flat
non-accelerating universes. The differences are normalised to be zero
at low redshifts. Error bars correspond to 68.3\,\% CLs.}
\end{center}
\end{figure}

For the non-accelerating case, $H\propto (1+z)$, and
\begin{equation}
  d_{c,n} = \frac{1}{H_0}\ln{(1+z)}\, ,
\end{equation}
and
\begin{equation}
  \label{eq:dln}
  d_{L,n} = \frac{1+z}{H_0\sqrt{-\ok}}\sin\left[ \sqrt{-\ok} \ln{(1+z)}\right] \, ,
\end{equation}
where subscript $n$ refers to non-accelerating expansion. For a flat,
non-accelerating, universe
\begin{equation}
  \label{eq:dlnf}
  d_{L,f} = \frac{1+z}{H_0}\ln (1+z)\, .
\end{equation}
For an empty, non-accelerating, universe, $\ok = 1$, and
\begin{equation}
  \label{eq:dlne}
  d_{L,e} = \frac{1+z}{H_0}\left(1+z- \frac{1}{1+z} \right)=\frac{z(z+2)}{H_0}\, .
\end{equation}
For small curvature ($\ok \ll 1$), defining $x=\sqrt{-\ok}d_c$, we can
use a Taylor expansion in $x$ and write
\begin{eqnarray}
  \frac{H_0 d_L}{1+z} &=& \frac{1}{\sqrt{-\ok}}\sin{(x)}\\
  &=& \frac{1}{\sqrt{-\ok}}\left(x-\frac{x^3}{6} \right)+\mathcal{O}(x^5)\\
  &\simeq & H_0\left(d_c + \frac{\ok}{6}d_c^3\right) \, . 
\end{eqnarray}
The difference between the observed SN Ia magnitudes and those
expected in a non-accelerating universe is given by
\begin{equation}\label{eq:deltam}
  \Delta m \equiv m-m_n = \frac{5}{\ln 10}\ln{\left(\frac{d_L}{d_{L,n}}\right)}\, .
\end{equation}
In \citet{2008JCAP...02..007S}, the inequality (valid in a flat universe)
\begin{eqnarray}
  \label{eqarr:dl}
  d_L(z)&=&\frac{(1+z)}{H_0}\int_{0}^{z}{\exp{\left[-\int_0^v
  \frac{[1+q(u)]du}{(1+u)}\right]}dv}\\
  &<&\frac{(1+z)}{H_0}\int_0^z\frac{dz}{(1+z)}=\frac{(1+z)}{H_0}\ln(1+z)\, ,
\end{eqnarray}
was considered as evidence ($\sim 5\sigma$ CL) for some period of
acceleration up to redshift $z$, i.e., $q<0$ at some
redshift. Allowing for spatial curvature, the evidence becomes
significantly weaker, or $1.8\sigma$.  We note [Eq.~(\ref{eq:dlne})],
that this is equivalent to studying the sign of $\Delta m$, if the
universe is flat. Assuming we can get rid of the dependence on $H_0$
and $M$ by normalising the difference to be zero at low redshifts, we
can study the dependence of the curvature term by looking at the
difference when subtracting empty and flat non-accelerating
cosmologies (Fig.~\ref{fig:magdiff}). It is obvious that $\Delta m$ in
fact is positive for $z\sim 0.5$, regardless of the curvature of the
universe.
%----------------------------------------------------------------------
%----------------------------------------------------------------------
A similar comparsion between SNe Ia at low and mid/high redshifts was
recently used in \citet{2008arXiv0810.4484S} to provide a
calibration-independent test of the accelerated expansion of the
universe, the conclusion being that the universe has accelerated at
{\em some} epoch at $\sim 4\sigma$ CL. From Eq.~(\ref{eqarr:dl}), it is
evident that the sign of $\Delta m$ only tells us whether the
integrated expansion up to $z$ is accelerating or decelerating on
average, {\em not} the state of acceleration at a given redshift.

\begin{figure}
\begin{center}
\includegraphics[angle=0,width=.65\textwidth]{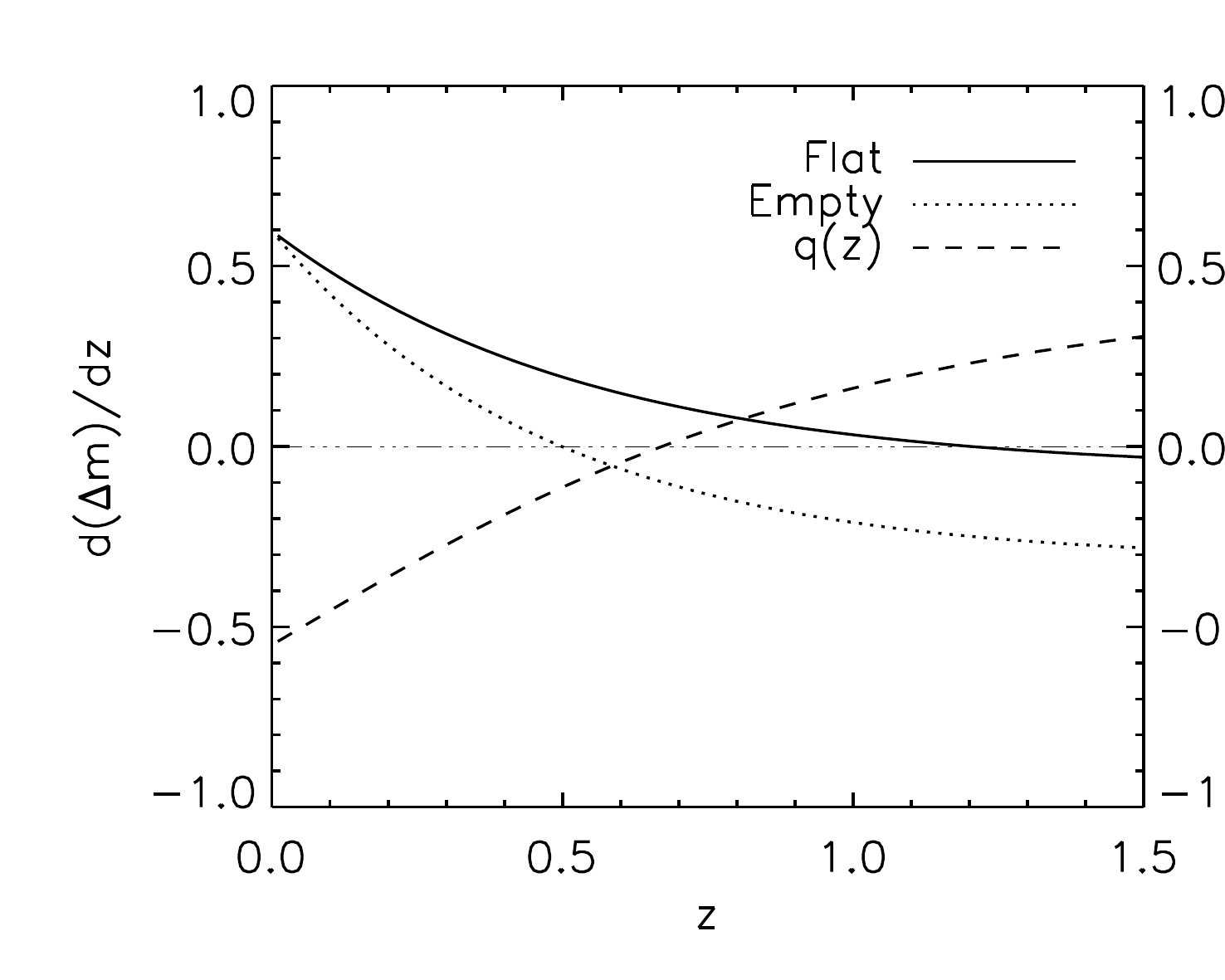}
\caption{\label{fig:acctestcon} $(\Delta m)'=d(\Delta m)/dz$ compared to $q(z)$ for 
the concordance model. The solid and dotted lines shows the derivative
of the difference between the concordance and the flat and empty
non-accelerating models, respectively. There is not a one-to-one
correspondence between the sign of $(\Delta m)'$ and the state of
acceleration.}
\end{center}
\end{figure}

We therefore turn to the second common claim, namely that if the
difference $\Delta m$ increases with redshift, the universe is
accelerating at that very redshift. 
The derivative of $\Delta m$ with respect to redshift is given by
\begin{equation}
  \label{eq:dmdz}
  (\Delta m)' = \frac{5}{\ln 10}\left(\frac{d_L'}{d_{L}} 
  -\frac{d_{L,n}'}{d_{L,n}}\right) \, .
\end{equation}
We note that $(\Delta m)'$ is independent of $H_0$ and $M$. 
For $(\Delta m)'>0$, we have
\begin{equation}
  \frac{d_L'}{d_{L}} >\frac{d_{L,n}'}{d_{L,n}}\, .
\end{equation}
Assuming a flat universe, this amounts to 
\begin{equation}
  H(z)\int_0^z \frac{dz}{H(z)}<(1+z)\ln (1+z)\, .
\end{equation}
From Eq.~(\ref{eq:dmdz}), it is evident that $(\Delta m)'$ does not
depend only on the state of acceleration at the given redshift, but
also on the integral of the expansion. The sign of $(\Delta m)'$ is
thus not directly related to the state of acceleration.  As a specific
example, consider the concordance model with $\om =0.3$ and $\ola =
0.7$ in Fig.~\ref{fig:acctestcon}. The lines corresponding to $(\Delta
m)'$ and the deceleration parameter does not cross the zero line at
the same redshift.
%In Fig.~\ref{fig:acctest}, we plot $(\Delta m)'$ as a
%function of redshift for different values of $\alpha$, assuming that
%$H(z)\propto (1+z)^\alpha$, where $\alpha = 0$ is the dividing line
%between acceleration and deceleration. In this case, we have a
%one-to-one correspondence between the sign of $(\Delta m)'$ and the
%deceleration parameter since they do not change sign over the redshift
%interval.
%----------------------------------------------------------------------
%----------------------------------------------------------------------
We conclude that visually judging the state of the universal
acceleration from a SN Ia Hubble diagram is non-trivial.

%%%%%%%%%%%%%%%%%%%%%%%%%%%%%%%%%%%%%%%%%%%%%%%%%%%%%%%%%%%%%%%%%%%%%%%
\subsection{Piecewise constant $q(z)$}\label{sec:piecewise}
We divide our redshift range in $[z_0=0,z_1,z_2,z_3,\ldots]$, where
between redshift, $z_{i-1}$ and $z_i$, we have a constant deceleration
parameter $q_i$.  In that bin, we then have $H(z)\propto
(1+z)^{q_i+1}$. Since the normalisation of the Hubble parameter is
marginalised over, we can put $H(z_0)=H_0=1$.  For $0<z<z_1$, we then
have $H(z)=(1+z)^{q_1+1}$, for $z_1<z<z_2$, we have
$H(z)=(1+z_1)^{q_1-q_2}(1+z)^{q_2+1}$, etc.

\begin{figure}
\begin{center}
\includegraphics[angle=0,width=.49\textwidth]{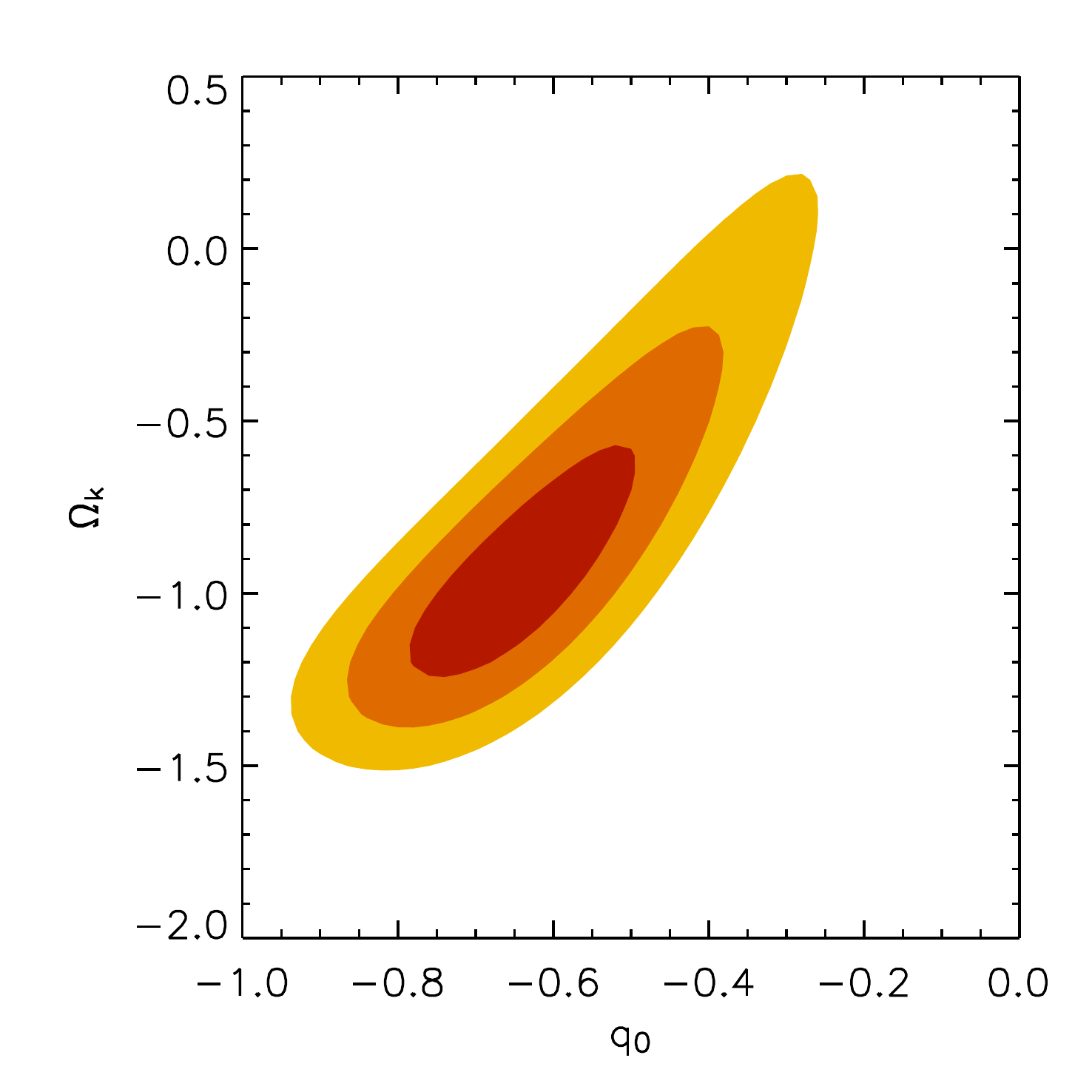}
\includegraphics[angle=0,width=.49\textwidth]{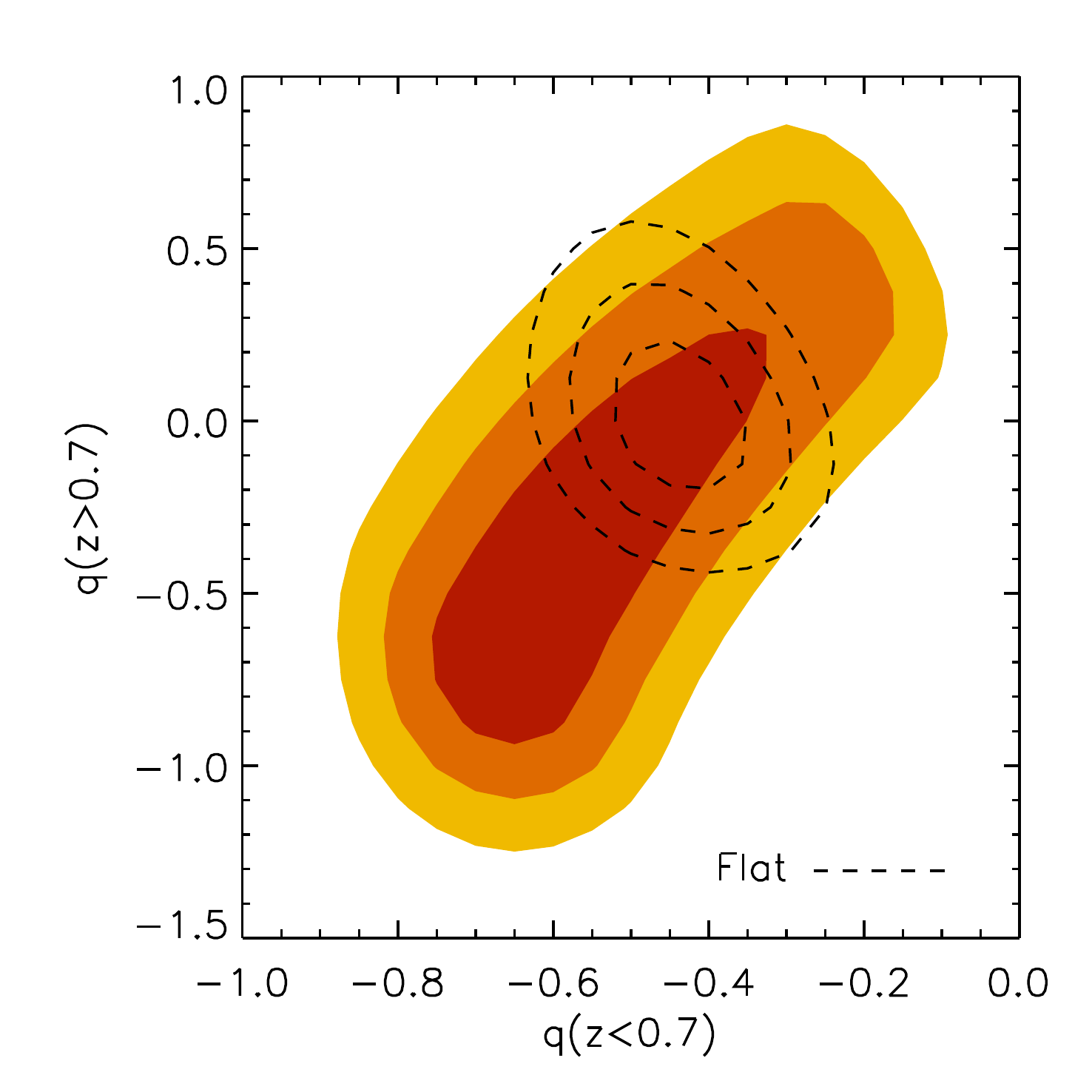}
\caption{\label{fig:qgridcomb} {\em Left panel:} SN Ia constraints on a model where 
the deceleration parameter has a constant value $q(z)=q_0$. {\em Right
panel:} SN Ia constraints on a model where the deceleration parameter
$q$ has one constant value at $z<0.7$ and one constant value at
$z>0.7$. Allowing for curvature increases the size of the contours by
a factor of $\sim 2$. The evidence for acceleration at $z<0.7$ is
$\sim 7\sigma$ and $\sim 5\sigma$, respectively. Contours correspond
to 68.3\,\%, 95.4\,\% and 99.7\,\% CLs.}
\end{center}
\end{figure}

The simplest case of a piecewise constant $q(z)$ would be to have a
single constant value of the deceleration parameter, i.e., $q(z)=q_0$
and $H(z)=H_0(1+z)^{q_0+1}$, at all times. Actually, such a simple
model gives a reasonable fit to SN Ia data, especially when allowing
for curvature, see left panel of Fig.~\ref{fig:qgridcomb}. To fit the
CMB and BAO data using this simple model however, requires a large
amount of fine-tuning since the quality of the fit depends very
sensitively on the exact values of $\ok$ and $q_0$. Also, since CMB
and BAO prefers a close to flat solution, the model provides a poor
fit to all data combined.

\begin{figure}
\begin{center}
\includegraphics[angle=0,width=.49\textwidth]{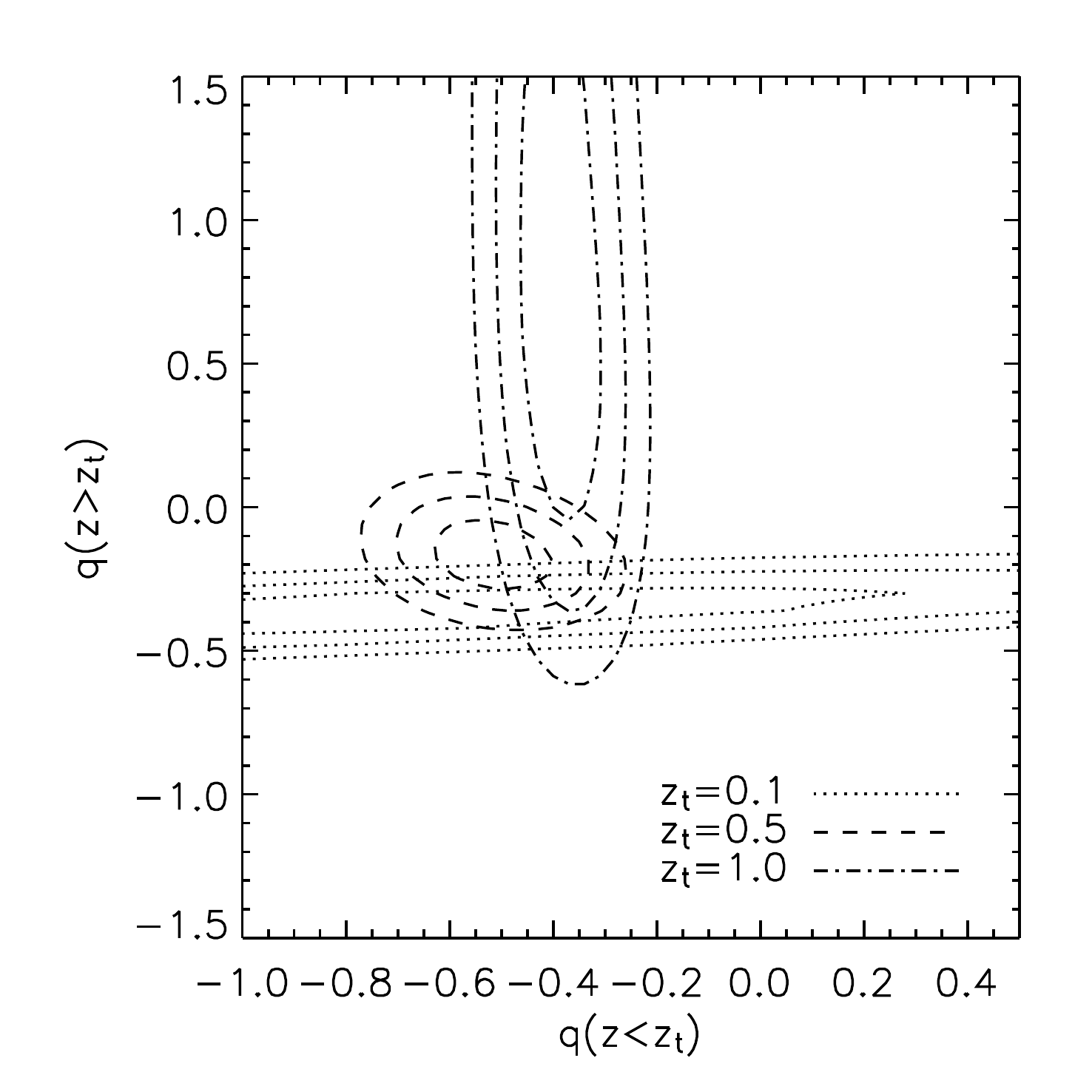}
\includegraphics[angle=0,width=.49\textwidth]{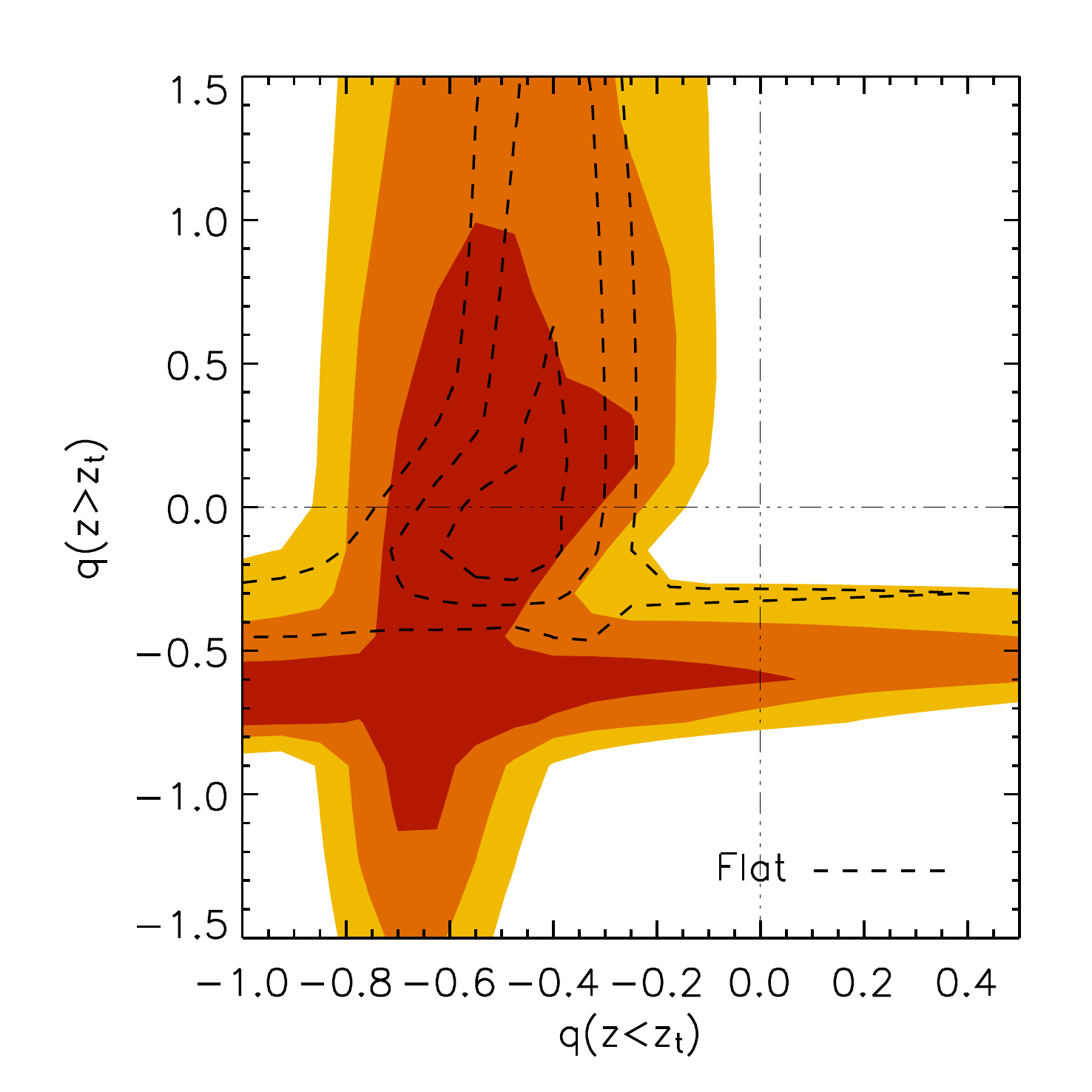}
\caption{\label{fig:ztflat} SN Ia constraints the deceleration parameter 
$q$ when varying the transition redshift $z_t$.  In the left panel,
confidence contours for three discrete values of $z_t$ in a flat
universe is shown. In the right panel, results when marginalising over
the transition redshift in the interval $0.1<z_t<1.0$ are shown, with
and without the assumption of a flat universe. Contours correspond to
68.3\,\%, 95.4\,\% and 99.7\,\% CLs.}
\end{center}
\end{figure}

\begin{figure}
\begin{center}
\includegraphics[angle=0,width=.49\textwidth]{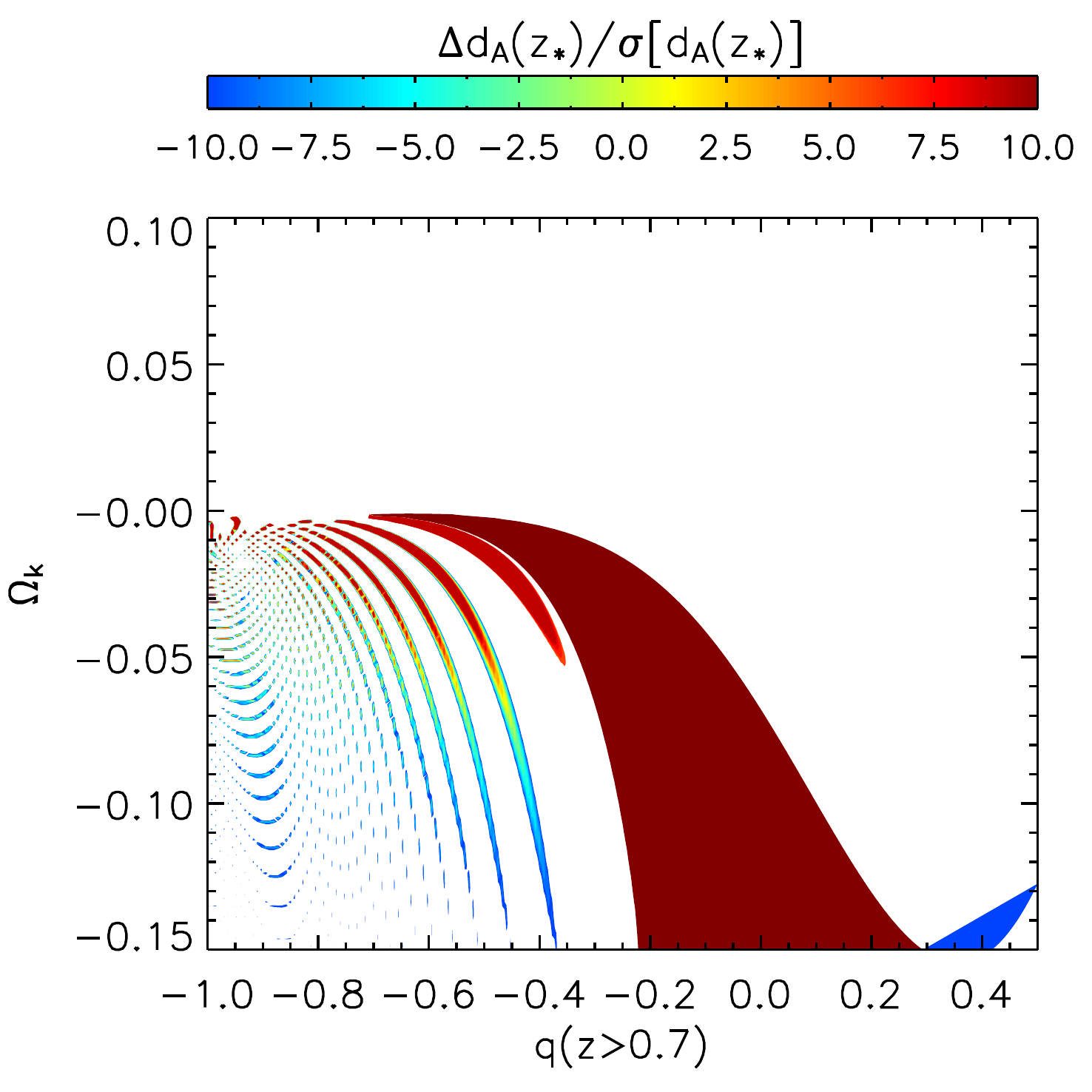}
\includegraphics[angle=0,width=.49\textwidth]{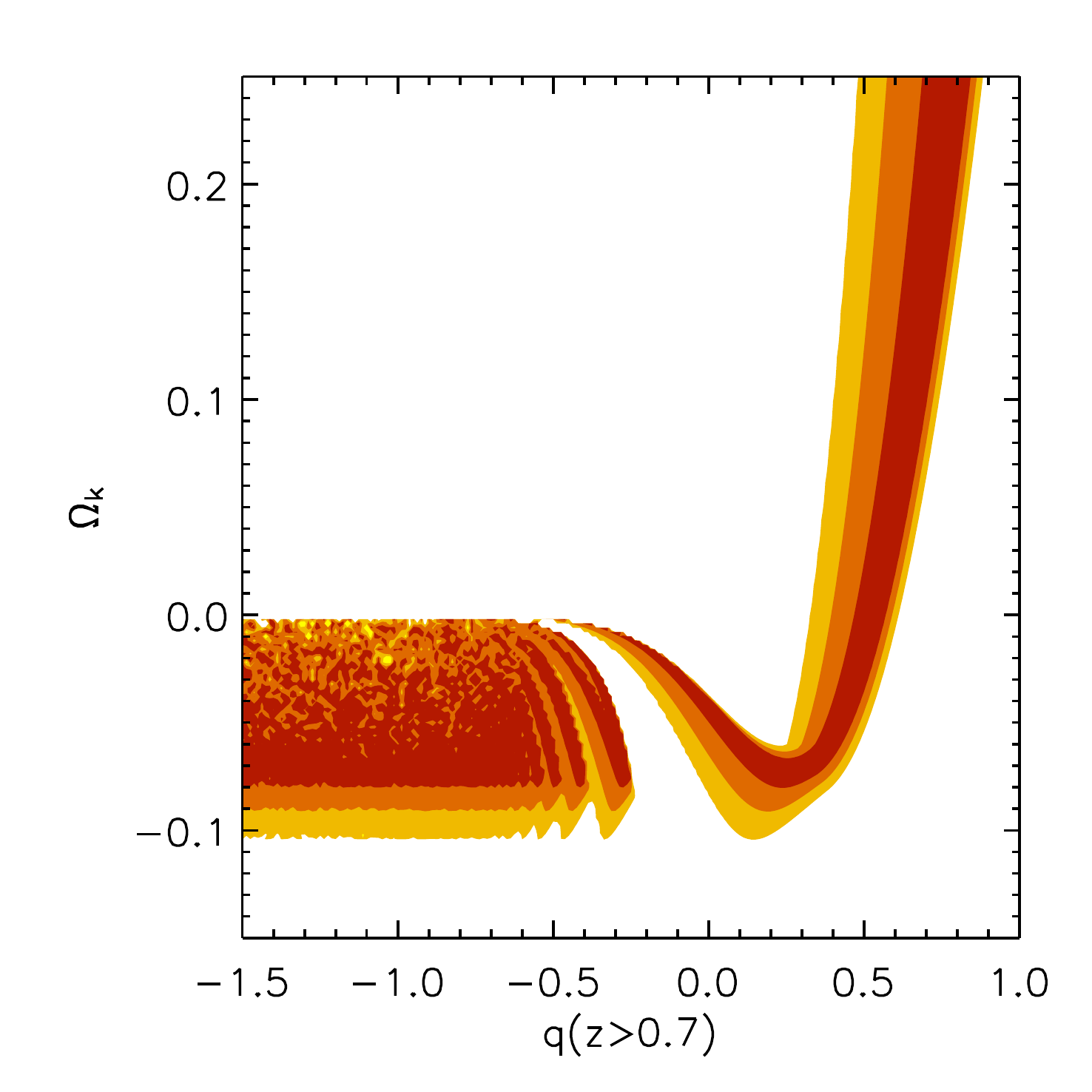}
\caption{\label{fig:qgridcmbbo2}
An example of the strong degeneracy between the curvature and the
deceleration parameter at high redshifts in the simple model where
$q(z)$ has one constant value at $z<0.7$ and one constant value at
$z>0.7$. In the left panel, the difference between the measured value
and theoretical values of $d_A(z_*)$ for a fixed value of
$D_V(z=0.2)$, normalised with the measured error on $d_A(z_*)$, is
shown. In the right panel, full confidence contours employing SN Ia,
CMB and BAO data are shown. The degeneracy is not sensitive to the
exact redshift binning of $q(z)$. The shaded areas represent 68.3\,\%,
95.4\,\% and 99.7\,\% CLs.}
\end{center}
\end{figure}

We next turn to an analysis along the lines of
\citet{2002ApJ...569...18T} where we assume that the deceleration
parameter $q(z)$ has one constant value at $z<z_t$ and one constant
value at $z>z_t$. We first choose $z_t = 0.7$ in order to maximise our
chances to detect a difference in the change of the expansion rate,
since this is the transition redshift for the concordance cosmology
that we know provides a good fit to the data. Results using SN Ia data
are shown in the right panel of Fig.~\ref{fig:qgridcomb}.
%----------------------------------------------------------------------
%----------------------------------------------------------------------
Allowing for curvature increases the size of the contours by a factor
of $\sim 2$. Incorporating also CMB and BAO data has the effect of
constraining the curvature down to the level where the combined
results are very similar to SN Ia results when assuming zero
curvature, except for a tail of very low values for $q(z>0.7)$ if the
universe has a small positive curvature ($\ok <0$, see
Fig.~\ref{fig:qgridcmbbo2}). Note that the data does not show any
evidence for deceleration at $z>0.7$. However, the evidence for
acceleration at $z<0.7$ is very strong; $\sim 5\sigma$ for SN Ia data
only and $\sim 7\sigma$ including CMB and BAO data or assuming a flat
universe. However, as shown in \citet{2006ApJ...649..563S},
marginalising over the transition redshift, considerably relaxes the
constraints on the expansion history. In the left panel of
Fig.~\ref{fig:ztflat}, we show how the confidence contours change with
the transition redshift, $z_t$, assuming a flat universe. For high
$z_t$, results are mostly sensitive to $q(z<z_t)$, and vice versa. In
the right panel of Fig.~\ref{fig:ztflat}, results when marginalising
over the transition redshift in the interval $0.1<z_t<1.0$ are shown,
with and without the assumption of a flat universe. The upper, right
quadrant, corresponding to expansion histories without any
acceleration, can be ruled with high confidence, independent of
curvature. However, all other possibilities are still viable,
including deceleration at low redshifts and acceleration at higher
redshifts, if the transition redshift is low enough ($z_t\sim 0.1$).
%----------------------------------------------------------------------
%----------------------------------------------------------------------

\begin{figure}
\begin{center}
\includegraphics[angle=0,width=.49\textwidth]{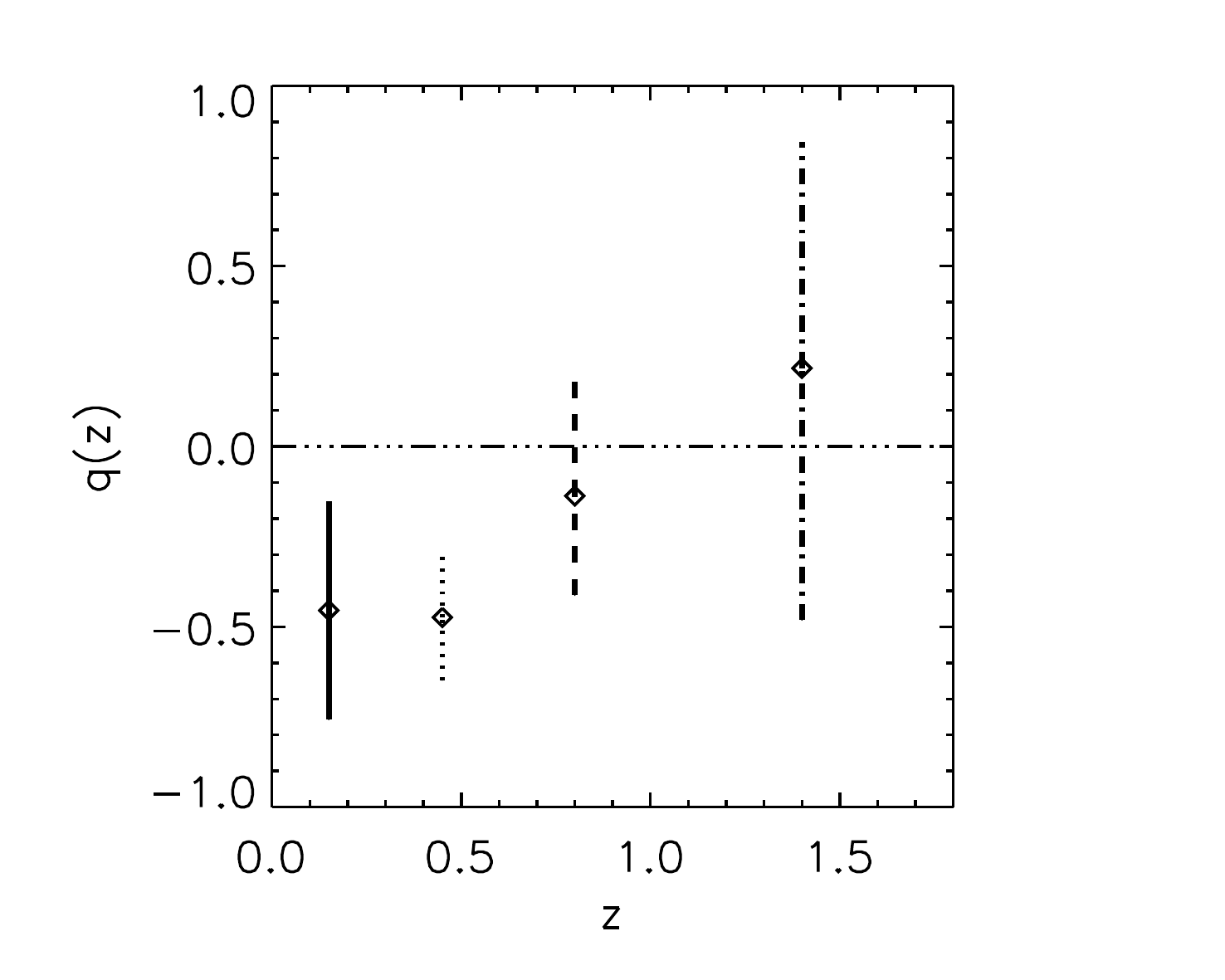}
\includegraphics[angle=0,width=.49\textwidth]{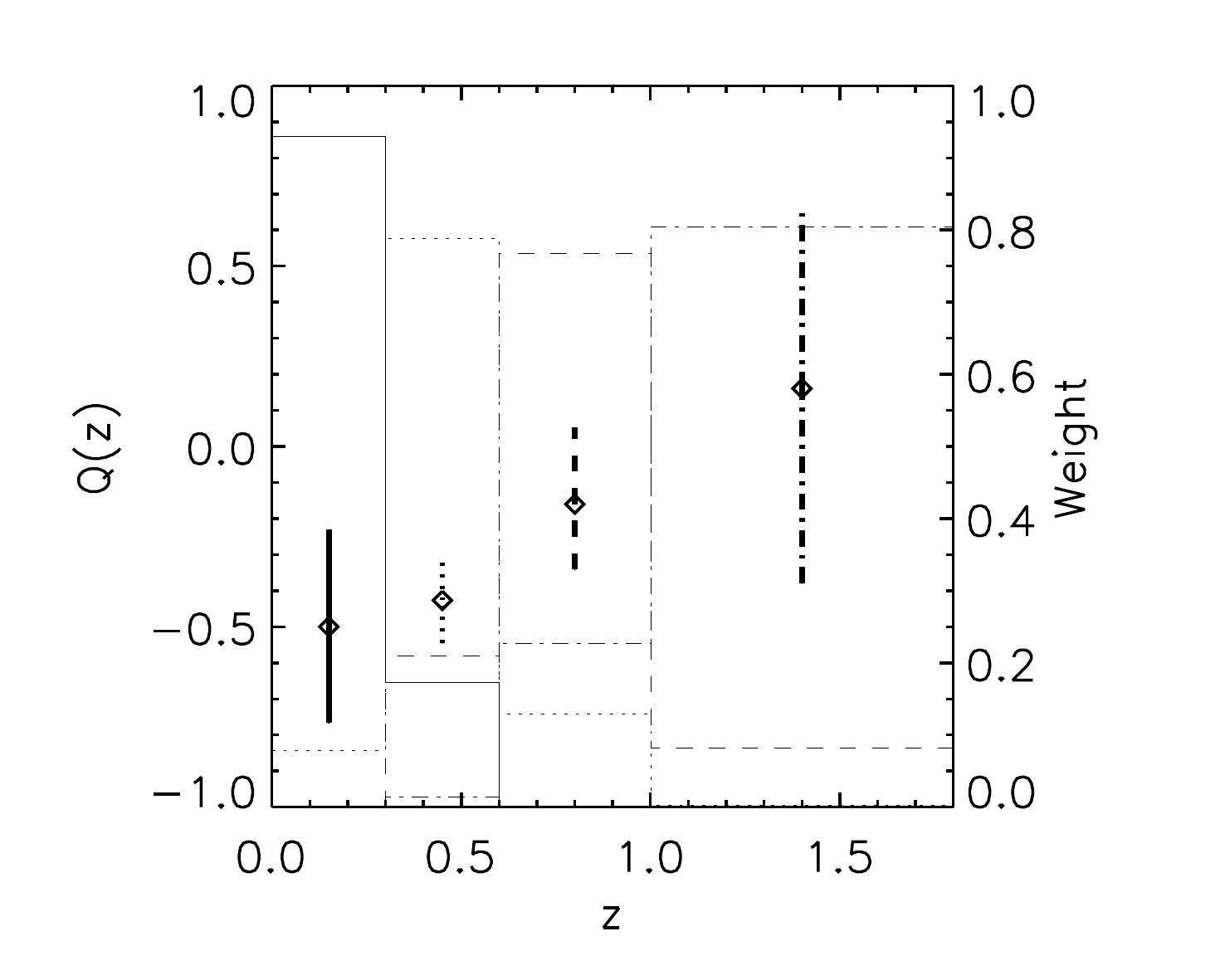}
\caption{\label{fig:snnok} SN Ia constraints on 
$q(z)$ in bins $z = [0.,0.3,0.6,1.,1.8]$, assuming a flat
universe. The left panel shows results for $q_i$, and the right panel
decorrelated $Q_i$ together with the corresponding weights. Error bars
represent 95.4\,\% CLs.}
\end{center}
\end{figure}

Next, we allow for a finer redshift resolution in $q(z)$. Note that
the $q_i$ are correlated, but we can decorrelate the $q_i$ estimates
by (following Ref.~\cite{2005PhRvD..71b3506H}) changing the basis
through an orthogonal matrix rotation that diagonalises the covariance
matrix. This corresponds to applying a weight function to the $q_i$ to
obtain decorrelated $Q_i$. These are linear combinations of $q_i$
where the weight function quantifies the redshift dependence of
$Q_i$. First, we use SN Ia data only to constrain $q(z)$ in bins $z =
[0.,0.3,0.6,1.,1.8]$, assuming a flat universe. Results for $q_i$ and
the decorrelated $Q_i$ together with the corresponding weights are
presented in Fig.~\ref{fig:snnok}. 
%----------------------------------------------------------------------
%----------------------------------------------------------------------
It is interesting to note that we again detect acceleration at low
redshift with high confidence, but that the data do not require
deceleration at higher redshifts. Allowing for curvature increases the
error bars on $q(z)$ somewhat, but does not change our qualitative
results.

\begin{figure}
\begin{center}
\includegraphics[angle=0,width=.49\textwidth]{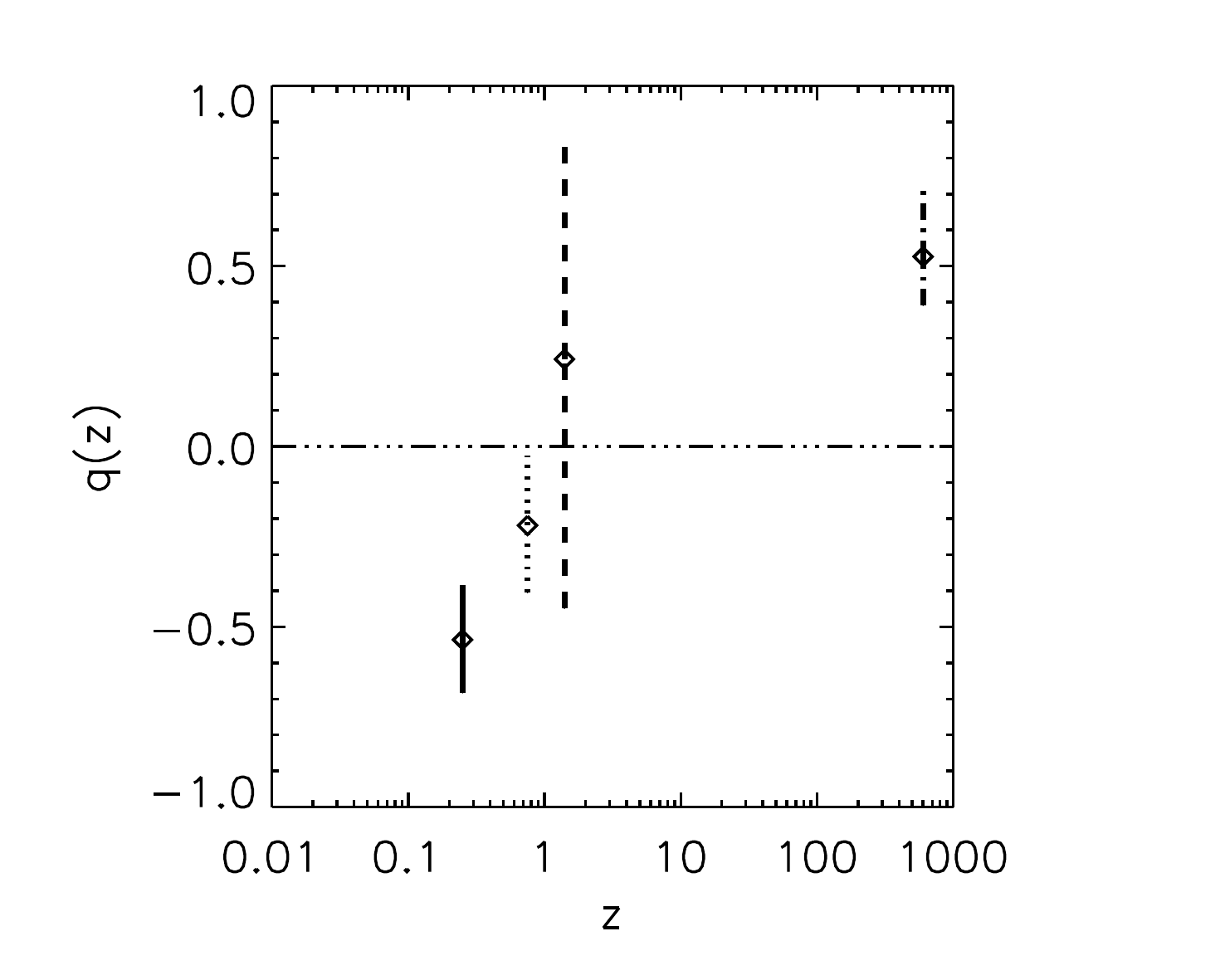}
\includegraphics[angle=0,width=.49\textwidth]{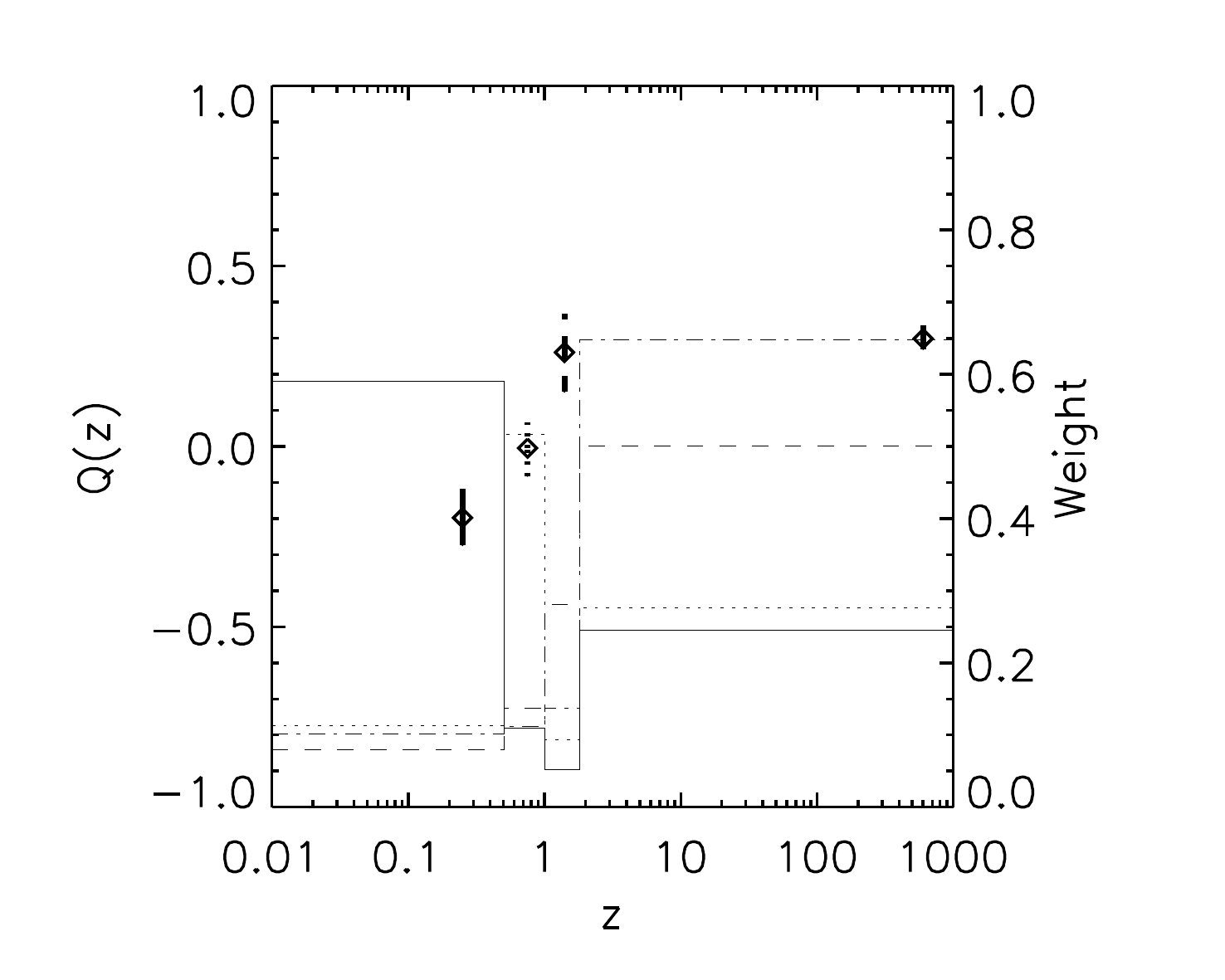}
\caption{\label{fig:sncmbbonok} 
Constraints on $q(z)$ in redshift bins $z = [0.,0.5,1.,1.8,1100.]$
combining SN Ia, CMB and BAO data, assuming a flat universe. The left
panel shows results for $q_i$, the right panel decorrelated $Q_i$ and
the corresonding weight functions. In the redshift interval
$1.8<z<1100$, the expansion is decelerating at high CL. Error bars
represent 95.4\,\% CLs. }
\end{center}
\end{figure}

Including CMB and BAO data, we can extend our analysis all the way out
to $z\sim 1100$ in redshift bins $z = [0.,0.5,1.,1.8,1100.]$. We can
see (Fig.~\ref{fig:sncmbbonok}) that in the redshift interval
$1.8<z<1100$, we in fact do see that the expansion was decelerating,
assuming a flat universe.
%----------------------------------------------------------------------
%----------------------------------------------------------------------
From the weight function, we can also see that the deceleration
parameter at different redshifts are very much correlated. If we relax
the assumption of flatness, our Monte Carlo Markov Chains fail to
converge, the reason being that the curvature and the deceleration
parameter at high redshifts exhibit very interesting degeneracies,
that also allow for solutions with high redshift acceleration and
positive curvature. This degeneracy is easy to understand when
considering the two component model where the deceleration parameter
$q(z)$ has one constant value at $z<z_t$ and one constant value at
$z>z_t$. For $z_t>0.35$, and a fixed $q(z<z_t)$, BAO constraints on
$D_V$ only has a very weak (sub-percent) dependence on curvature, and
the ratio between the scale of the BAO and CMB solely depends on the
angular diameter distance to $z_*\sim 1090$,
\begin{equation}\label{eq:deg}
  d_A(z_*)(1+z*) = \frac{1}{H_0\sqrt{-\ok}}\sin{\left[ \sqrt{-\ok}
  H_0 d_c(z_*)\right]} \, .
\end{equation}
In the left panel of Fig.~\ref{fig:qgridcmbbo2}, we have plotted the
difference between the measured and theoretical values of $d_A(z_*)$
for a fixed value of $D_V(z=0.2)$, normalised with the error on
$d_A(z_*)$. The characteristic degeneracy structure arises because of
the sinusoidal form of the angular distance in Eq.~(\ref{eq:deg}) and
allows for models with high redshift acceleration and positive
curvature to fit the data. In the right panel, the full confidence
contours for SN Ia, CMB and BAO data are shown. It is evident that
negative values for $q(z>0.7)$, i.e., corresponding to acceleration,
are allowed if the universe has a small positive curvature. We note
however, that such solutions require a considerable amount of fine
tuning of the deceleration and curvature parameters. 
%----------------------------------------------------------------------
%----------------------------------------------------------------------

%%%%%%%%%%%%%%%%%%%%%%%%%%%%%%%%%%%%%%%%%%%%%%%%%%%%%%%%%%%%%%%%%%%%%%%
\subsection{Parameterising $q(z)$}
There are many ways to parameterise $q(z)$. Clearly, a piecewise
constant function as used above only gives a limited amount of
information about $q(z)$; its average value over a certain redshift
range. Ideally, we would prefer something which can closely mimic what
the `true' $q(z)$ might be up to.  As an example, let us consider
\begin{equation}\label{eq:=qparam}
q(a)=q_0+q_a(1-a)=q_0+q_a\frac{z}{1+z}\, ,
\end{equation}
adopted from one of the most common parameterisation of the dark
energy equation of state, $w(z)$ \cite{2001IJMPD..10..213C}.  At zero
redshift, $q(a=1)=q_0$ and in the infinite past, $q(a=0)=q_0+q_a$.
This parameterisation appears to be reasonably flexible in the sense
that performing a least squares fit to many random dark energy models,
always provides an acceptable fit.
%----------------------------------------------------------------------
\begin{figure}
\begin{center}
\includegraphics[angle=0,width=.49\textwidth]{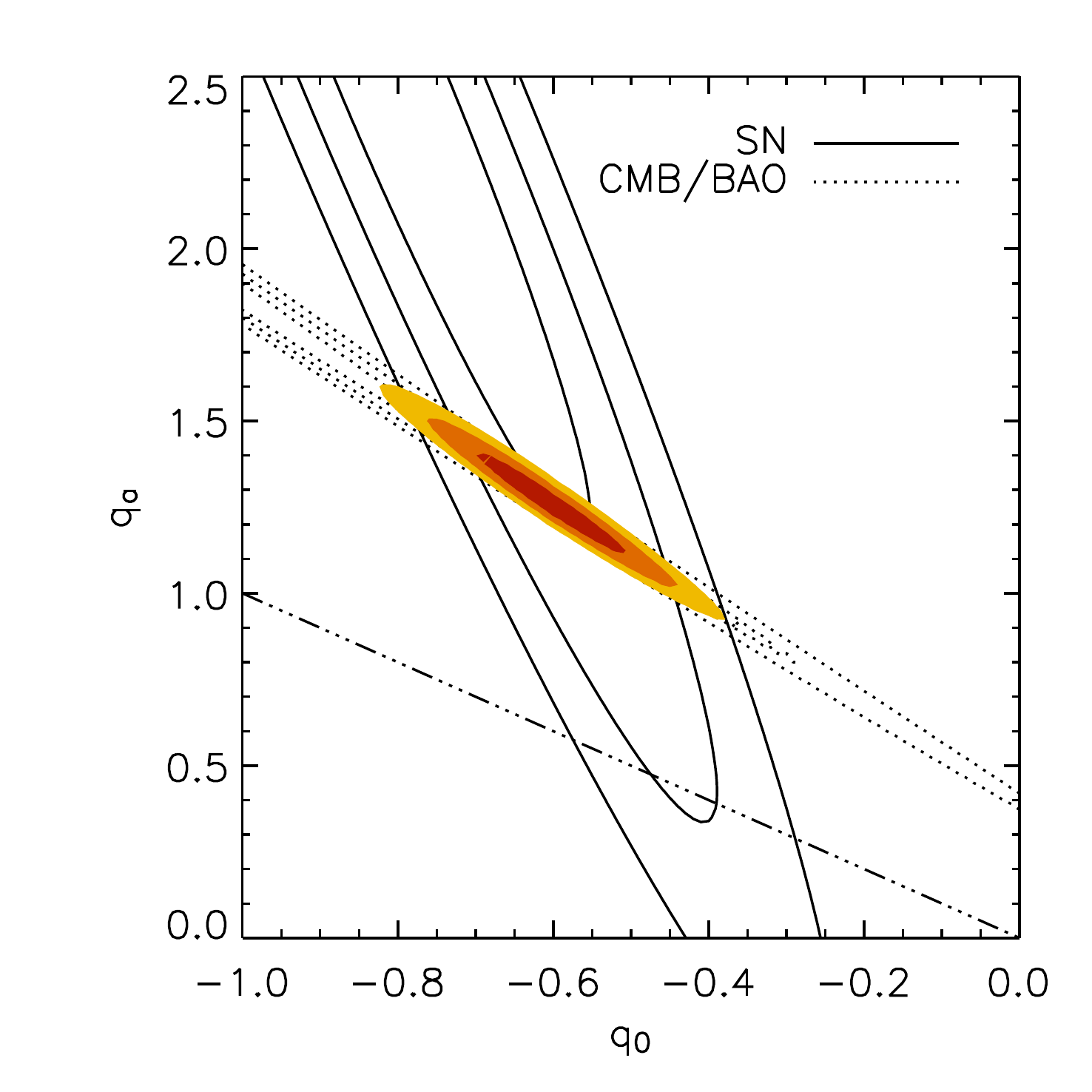}
\includegraphics[angle=0,width=.49\textwidth]{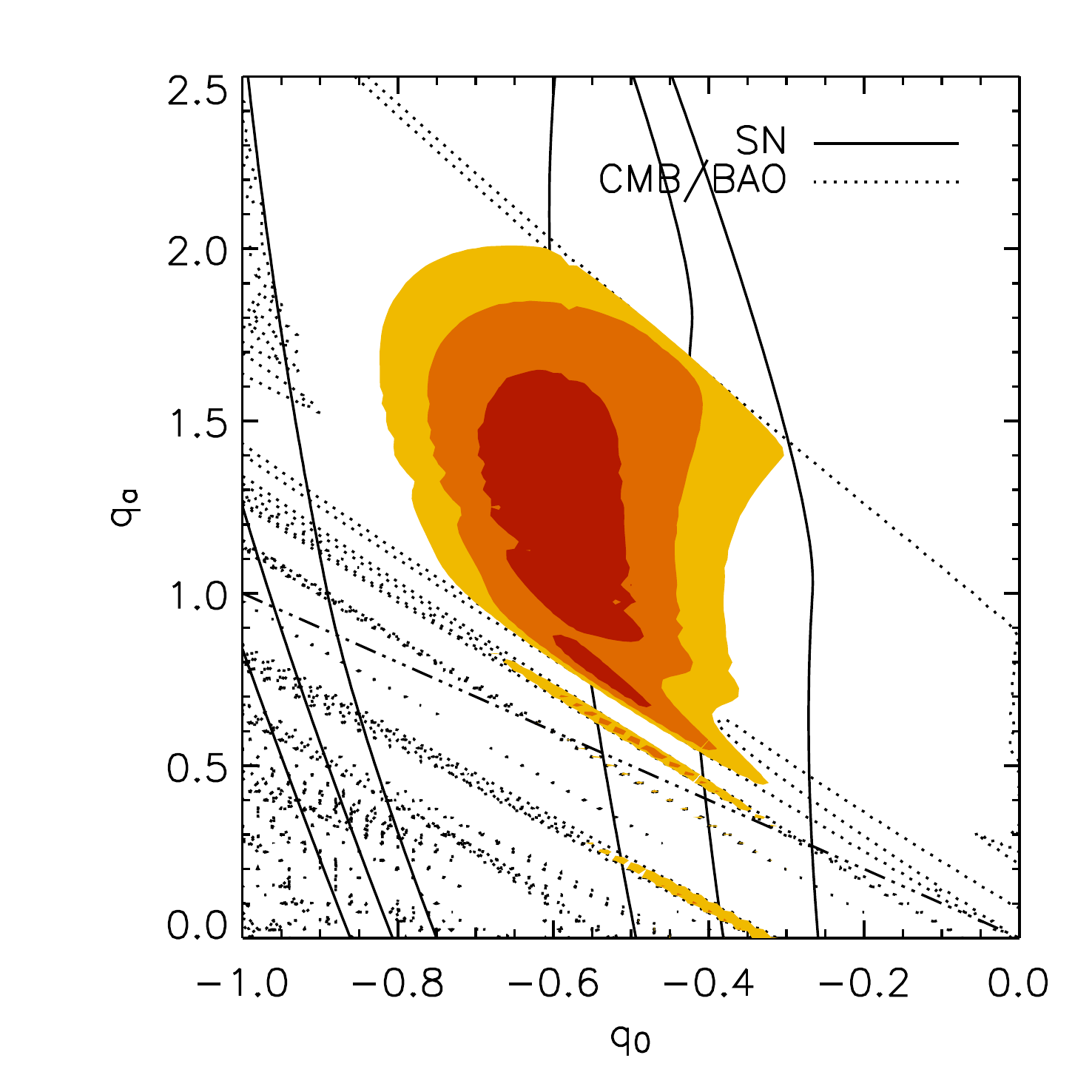}
\caption{\label{fig:q0qacomb} Constraints on $q_0, q_a$ for the 
parameterisation given by Eq.~(\ref{eq:=qparam}), assuming a flat
universe (left) and allowing for curvature (right). The dash-dotted
line is given by $q_0+q_a=0$, and separates regions with eternal
acceleration (lower left) from regions with past deceleration (upper
right). Using only SN Ia data, $q_0<0$ at $>5\sigma$ CL, irrespective
of the spatial curvature. The degenerate structures in the right panel
are discussed in Sec.~\ref{sec:piecewise}. Regions represent 68.3\,\%,
95.4\,\% and 99.7\,\% CLs.}
\end{center}
\end{figure}
%----------------------------------------------------------------------
In Fig.~\ref{fig:q0qacomb}, we show results for this parameterisation,
both with and without curvature. Once again curvature degrades all
constraints, though not as significantly as for other models for the
SN Ia constraints on $q(z)$. Therefore, fitting SN Ia data alone using
this parameterisation, we are able to conclude that the universe is
accelerating today at $>5\sigma$, irrespective of curvature.

For $q_a=0$, the model has $q(z)=q_0$ and is equivalent to the model
presented in Fig.~\ref{fig:qgridcomb}. Studying the contours at
$q_a=0$ in Fig.~\ref{fig:q0qacomb}, it is again obvious that this
simple model does not provide a good fit to the data, unless we allow
for curvature and an extreme fine-tuning of the parameters of the
model.

%%%%%%%%%%%%%%%%%%%%%%%%%%%%%%%%%%%%%%%%%%%%%%%%%%%%%%%%%%%%%%%%%%%%%%%
\subsection{The Hubble parameter}
\begin{figure}
\begin{center}
\includegraphics[angle=0,width=.49\textwidth]{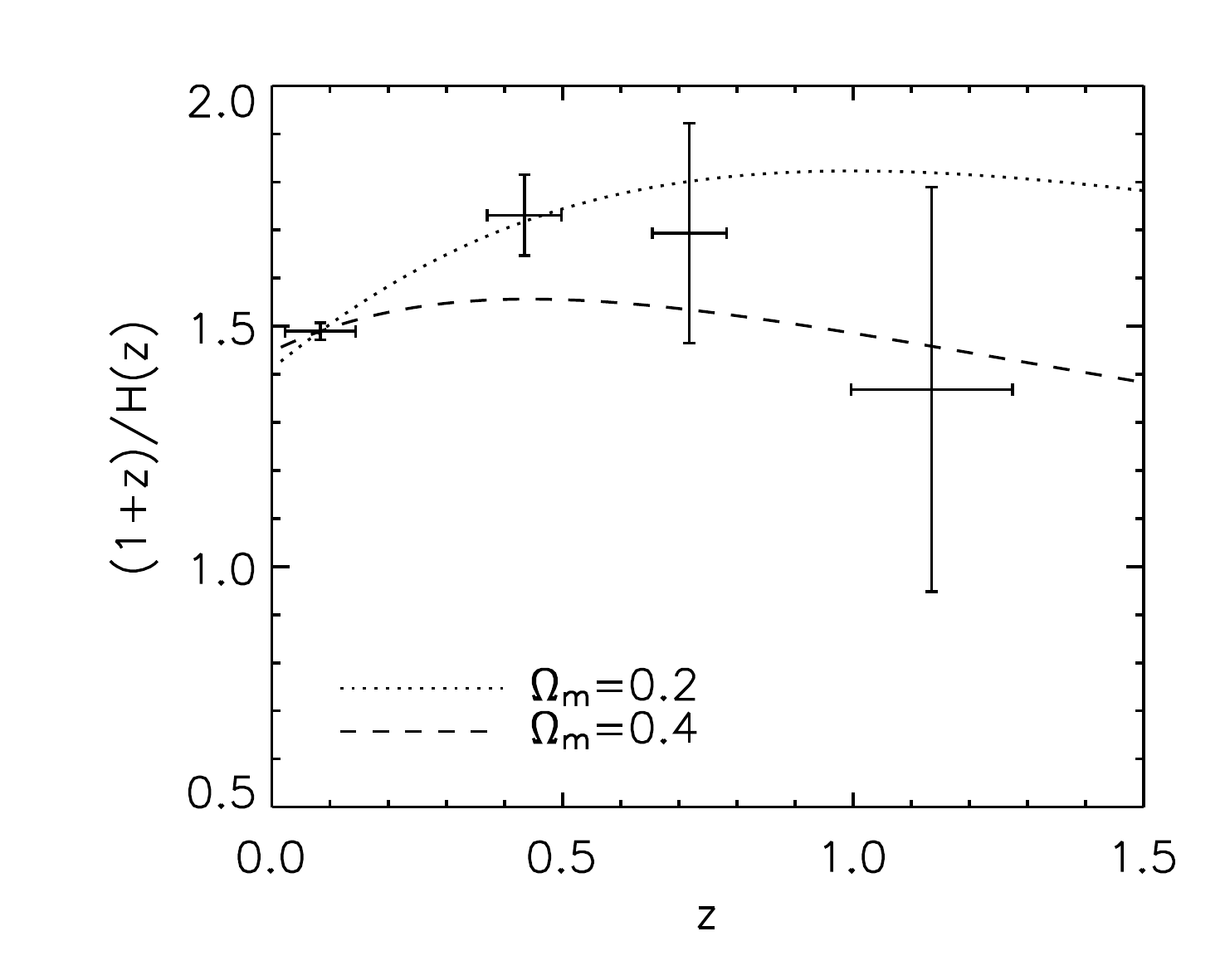}
\includegraphics[angle=0,width=.49\textwidth]{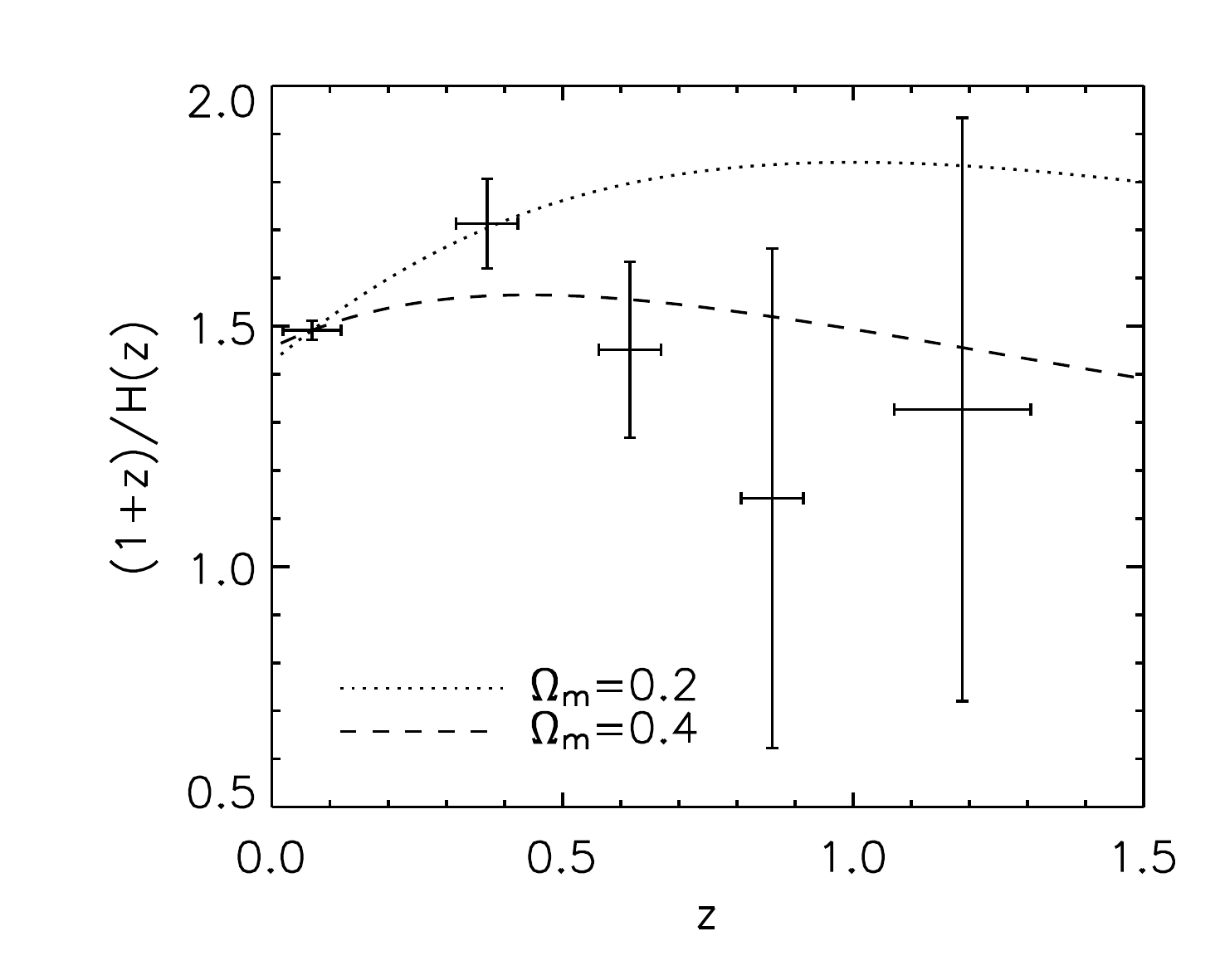}
\caption{\label{fig:SCPUnion} SN Ia constraints on $\dot
a^{-1}=(1+z)/H(z)$. Values increasing with redshift correspond to
acceleration and vice versa. The dotted lines correspond to a flat
universe with $\om = 0.2$ and $\om = 0.4$ from top to bottom. The
normalisation is arbitrary and is set to agree at low redshifts. In
the left panel of Fig~\ref{fig:SCPUnion}, redshifts bins of size
$\Delta z=0.3$ has been used whereas in the right panel, $\Delta
z=0.25$. Vertical error bars correspond to 68.3\,\% CLs. Acceleration
at low redshifts is clearly detected whereas deceleration at higher
redshifts is not.}
\end{center}
\end{figure}

Since it is possible, although difficult (see Fig.~\ref{fig:magdiff}),
to obtain the Hubble parameter by differentiating distances, which
would give us $q(z)$ directly, we can try to compare $\dot a^{-1}=(1+z)/H(z)$ at
different redshifts \cite{2007ApJ...659...98R}. We follow the
technique proposed in Wang and Tegmark \cite{2005PhRvD..71j3513W} to
extract the expansion history in uncorrelated redshift bins from SN Ia
data. The results are shown in Fig.~\ref{fig:SCPUnion}, where
increasing values of $(1+z)/H(z)$ corresponds to acceleration and vice
versa. Unfortunately, because of the differentiation of sparse and
noisy data, results are quite sensitive to the employed binning of the
data, especially at high redshift. In the left panel of
Fig~\ref{fig:SCPUnion}, redshifts bins of size $\Delta z=0.3$ has been
used whereas in the right panel, $\Delta z=0.25$. It is therefore
difficult to assess the exact CL for the trend seen in
Fig.~\ref{fig:SCPUnion} of increasing $(1+z)/H(z)$ -- corresponding to
acceleration -- at low redshifts and decreasing $(1+z)/H(z)$ --
corresponding to deceleration -- at high redshifts.  However,
independent of the bin size, acceleration at low redshifts seems to be
inevitable whereas deceleration at higher redshifts is not.
%----------------------------------------------------------------------
%----------------------------------------------------------------------

%%%%%%%%%%%%%%%%%%%%%%%%%%%%%%%%%%%%%%%%%%%%%%%%%%%%%%%%%%%%%%%%%%%%%%%
\subsection{Sliding Taylor expansion}
\begin{figure}
\begin{center}
\includegraphics[angle=0,width=.49\textwidth]{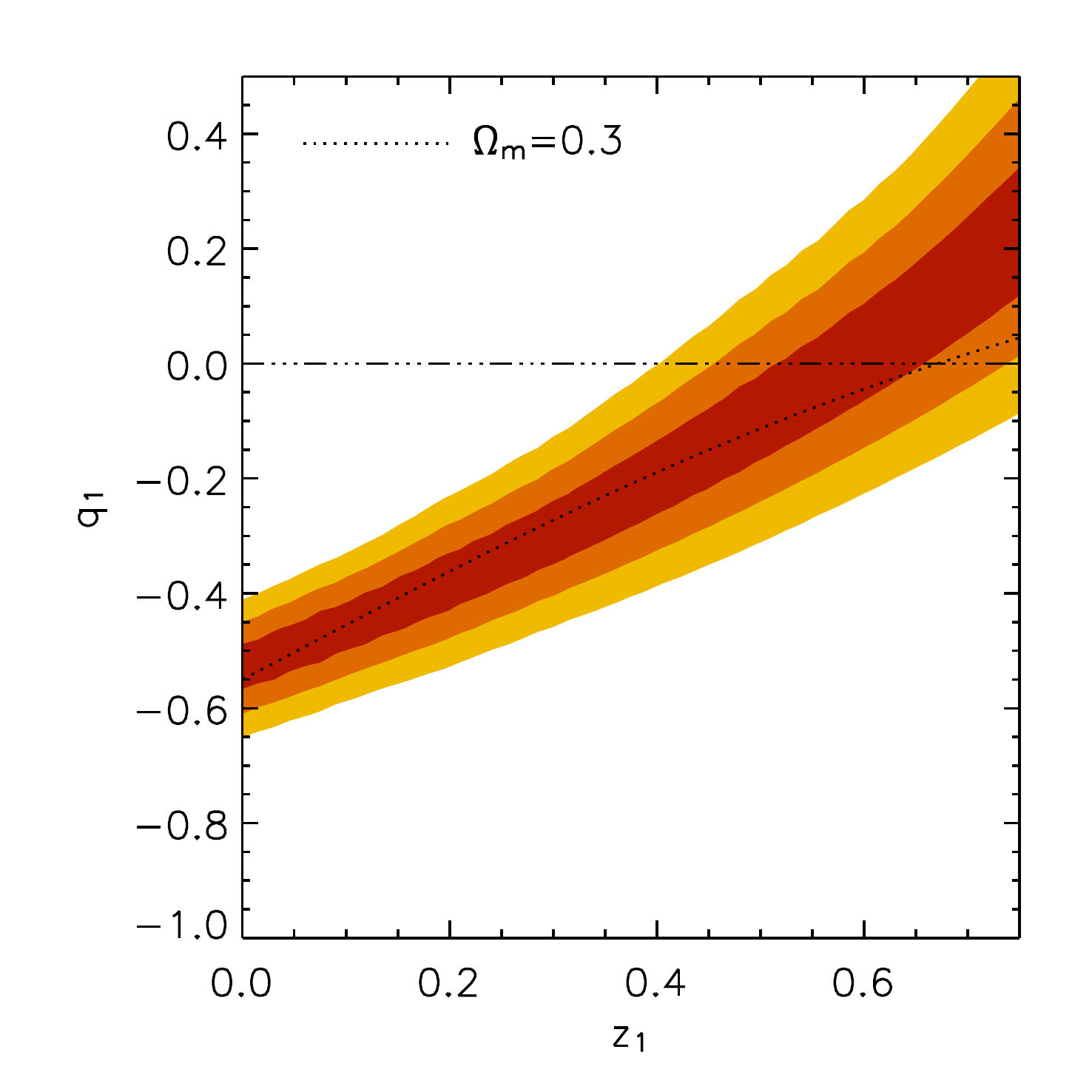}
\caption{\label{fig:sntest} Fitting the deceleration parameter, 
$q(z_1)=q_1$, with a sliding expansion redshift for data simulated within a flat,
cosmological constant universe with $\om = 0.3$. The bias between the
fitted $q_1$ and the true $q(z)$ (dotted line) caused by higher order terms
in the Taylor expansion lies within the 68.3\,\% CL for $z_1\lesssim
0.6$.}
\end{center}
\end{figure}

We can Taylor expand the scale factor around the current value
according to
\begin{eqnarray}
  a(t) &=& a_0 + \dot a_0 (t-t_0) + \frac{1}{2} \ddot a_0 (t-t_0)^2 + 
  \mathcal{O}(t-t_0)^3	\\
  &=& 1 + H_0 (t-t_0) - \frac{1}{2}q_0 H_0^2(t-t_0)^2 + \mathcal{O}(t-t_0)^3\, .
\end{eqnarray}
Since the comoving coordinate distance is given by
\begin{equation}
  d_c(t_0) = \int_{t_e}^{t_0}{\frac{dt}{a(t)}}\, ,
\end{equation}
we can write
\begin{equation}
  \label{eq:taylor}
  d_c(z) \simeq \frac{z}{H_0}\left[1-\frac{1+q_0}{2}z\right] \, .
\end{equation}
To first order, the distance is given by the expansion rate of the
universe today, or $H_0$, and to second order by the change in the
expansion rate today, or $q_0$. We now generalise
Eq.~(\ref{eq:taylor}) to allow for a Taylor expansion around any time,
or equivalently, redshift according to
\begin{eqnarray}
  a(t) &\simeq& a(t_1) + \dot a_1 (t-t_1) + \frac{1}{2} \ddot a_1 (t-t_1)^2\\
  &=& a_1\left[1+H_1 (t-t_1) - \frac{1}{2}q_1 H_1^2(t-t_1)^2\right]\, 
\end{eqnarray}
The comoving coordinate distance is given by
\begin{equation}
  d_c(t_0) \simeq \frac{1}{a_1}\left[(t_0-t_1)-\frac{H_1}{2}(t_0-t_1)^2-
  (t_e-t_1)+\frac{H_1}{2}(t_e-t_1)^2 \right] \, ,
\end{equation}
or, in terms of observables,
\begin{equation}
  d_c(z) \simeq
  \frac{z}{H_1}\left[1+\frac{z_1(1+q_1)}{1+z_1}-\frac{1}{2}\frac{(1+q_1)}{(1+z_1)}z\right]
  \, .
\end{equation}
We can now fit the parameter $q_1$, using SN Ia data for a sliding
expansion redshift $z_1$, while marginalising over $H_1$. In
Fig.~\ref{fig:sntest}, we show results obtained using simulated data,
corresponding to the Union08 data set, within the concordance
cosmology. We only include SNe for which $|z-z_1| < 0.75$ to guarantee
convergence, which is why this and subsequent plots only extend to $z
= 0.75$. As can be seen in Fig.~\ref{fig:sntest}, the bias from higher
order terms is within the 68.3\,\% CL for $z\lesssim 0.6$ and within
95.4\,\% CL all the way up to $z=0.75$.
%----------------------------------------------------------------------
%----------------------------------------------------------------------

\begin{figure}
\begin{center}
\includegraphics[angle=0,width=.49\textwidth]{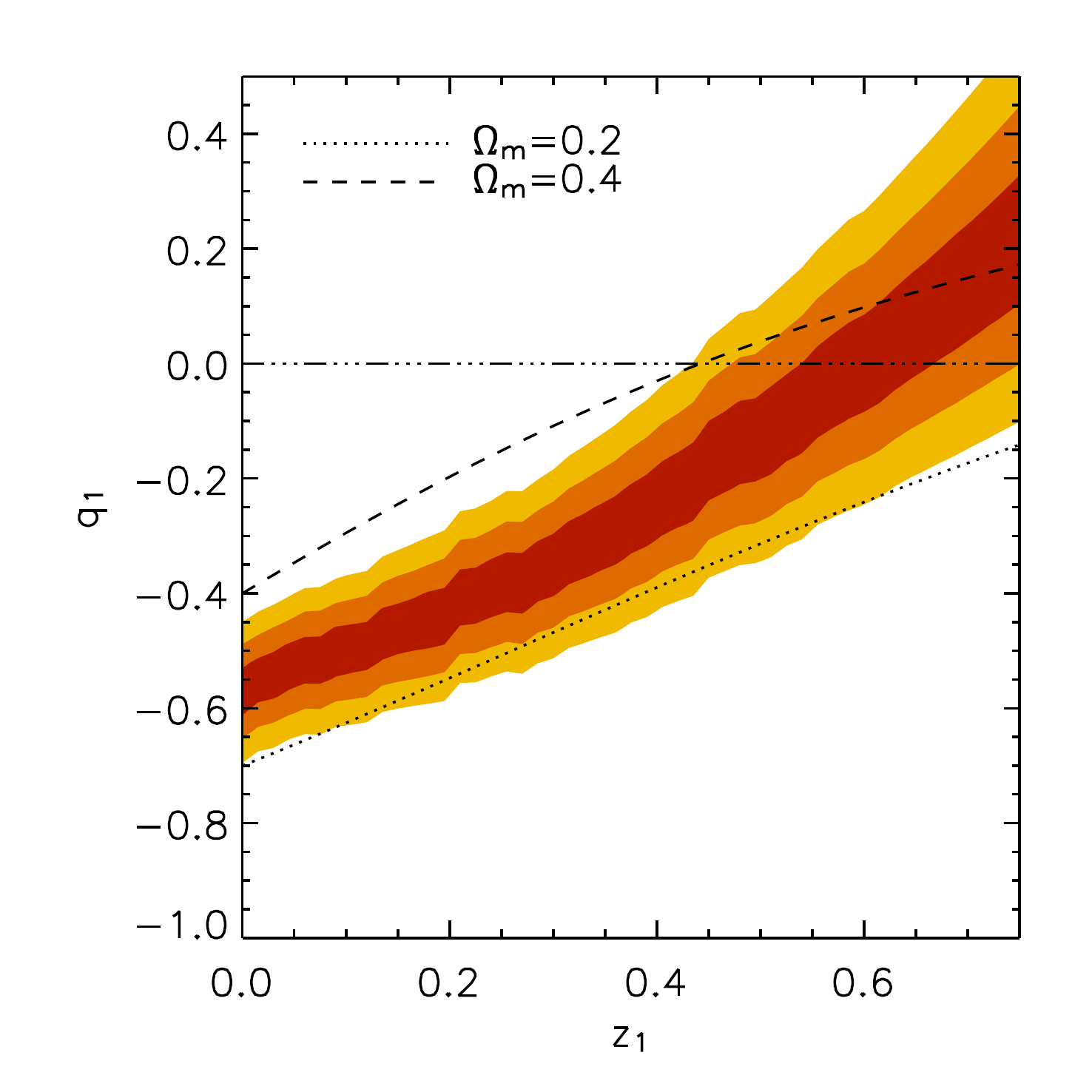}
\includegraphics[angle=0,width=.49\textwidth]{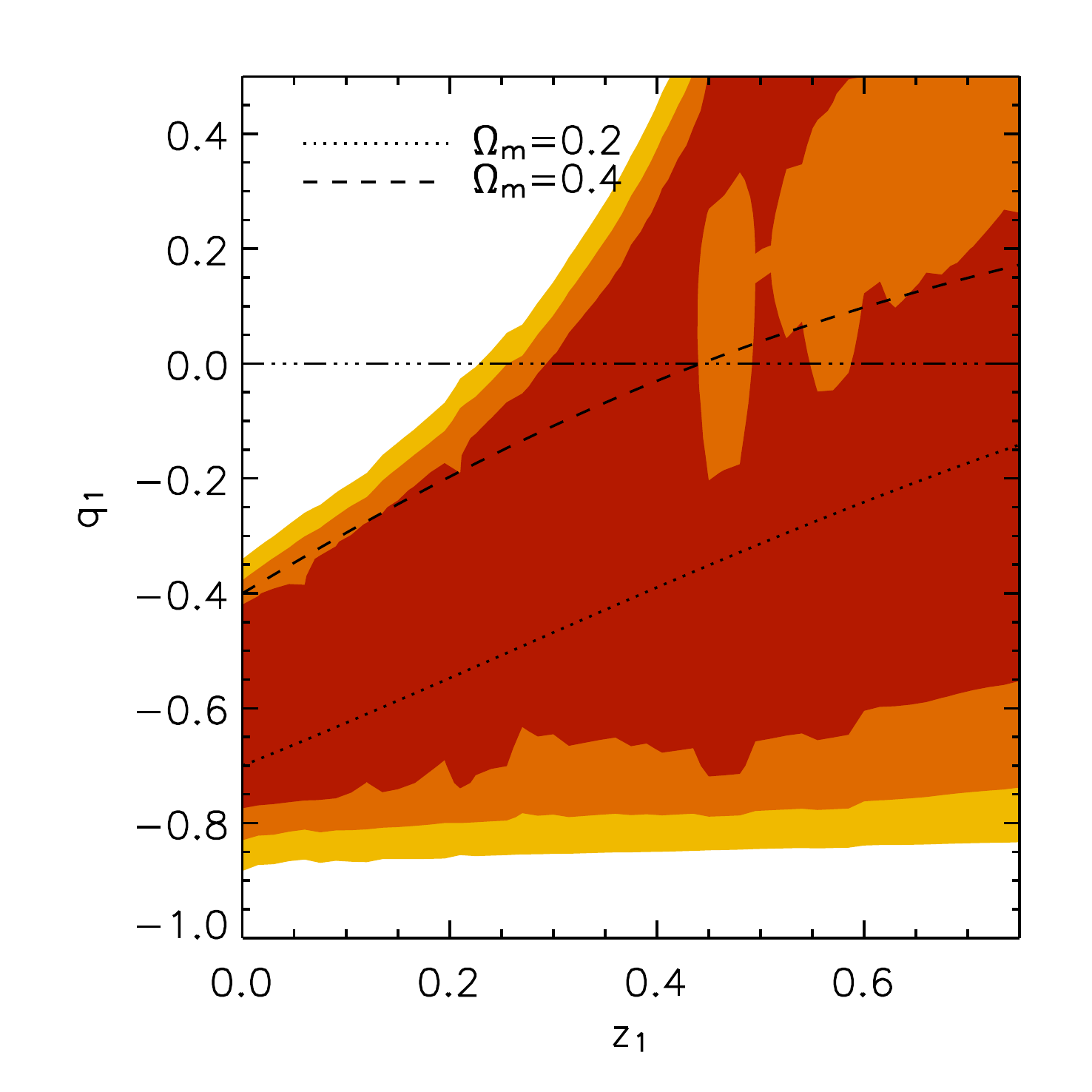}
\caption{\label{fig:taylor} Fitting the deceleration parameter, 
$q(z_1)=q_1$, with a sliding expansion redshift.  
Note that constraints on $q_1$ are
significantly looser when allowing for curvature (in this case
$-1<\ok<1$). However, the evidence for acceleration at $z=0$ is still
$>12\sigma$, including curvature. For a flat universe,
$q_1(z_1=0)<0$ at $\sim 15\sigma$ CL.  The lines correspond to the flat,
cosmological constant universes with $\om = 0.2$ and 0.4,
respectively. The shaded areas correspond to 68.3\,\%, 95.4\,\% and
99.7\,\% CLs.}
\end{center}
\end{figure}

In Fig.~\ref{fig:taylor}, results for the real Union08 data set, with
and without curvature ($-1<\ok<1$), is shown. Positive $\ok$ pushes
$q(z)$ towards more positive values and vice cersa. Again, we
only include data for which $|z-z_1| < 0.75$ to reduce bias from higher
order terms in the Taylor expansion.

Regardless of the expansion redshift, the sliding Taylor expansion is
able to provide a very good fit to the SN Ia data. Note however that
the results for $q_1$ at different redshifts are not independent, and
it is therefore not possible to interpret the results as giving a full
functional form of deceleration parameter $q(z)$. Assuming zero curvature,
the universe can be shown to be accelerating at $z\lesssim 0.5$ and
there is weak (95.4\,\%) evidence for decelaration at high
redshifts. Including curvature severly degrades these constraints,
especially the behaviour of $q(z)$ at high redshifts. Nevertheless,
using the sliding Taylor expansion, the evidence for acceleration at
$z=0$ is $>12\sigma$, even when allowing for curvature.

The reason for the closeness of the 2 and $3\sigma$ contours in the
right panel, is that at each redshift, the confidence contours in the
$[\ok,q_1]$-plane has a banana like shape. This has the effect that
the resulting confidence levels for $q_1$, which are obtained by
projecting the banana shaped contours on the $q_1$-axis, has a
non-gaussian shape.
%----------------------------------------------------------------------
%----------------------------------------------------------------------

%%%%%%%%%%%%%%%%%%%%%%%%%%%%%%%%%%%%%%%%%%%%%%%%%%%%%%%%%%%%%%%%%%%%%%%
%%%%%%%%%%%%%%%%%%%%%%%%%%%%%%%%%%%%%%%%%%%%%%%%%%%%%%%%%%%%%%%%%%%%%%%
\section{Summary}\label{sec:summary}
%%%%%%%%%%%%%%%%%%%%%%%%%%%%%%%%%%%%%%%%%%%%%%%%%%%%%%%%%%%%%%%%%%%%%%%
In this paper, we have investigated to what extent we can measure the
change in the universal expansion rate, without making any assumptions
about the energy content of the universe. Consequently we are limited
to geometrical data as opposed to methods that are sensitive to the
growth of structure in the universe. The data employed in this paper
includes the redshift-distance relation of Type Ia SNe, as well as the
the ratio of the scale of the baryon acoustic oscillations as
imprinted in the cosmic microwave background and in the large scale
distribution of galaxies. We have used several different methods to
constrain the expansion history of the universe, all of which give the
same qualitative result. 

From SN Ia data alone, it is evident that the universal expansion is
accelerating at low redshifts. In particular our new sliding expansion
redshift method is able to detect acceleration today at $>12\sigma$,
even allowing for curvature. Although there are
hints from SN Ia data that the universal expansion may have
decelerated at high redshifts~-- as expected in the concordance
cosmological model~-- we can only draw that conclusion with high
confidence if we also include CMB and BAO data, together with the
assumption of a flat or open universe. If the universe has a small
positive curvature ($\ok <0$), it is possible, although it requires a
certain level of fine tuning, to accommodate the data with
acceleration also at high redshifts.

%%%%%%%%%%%%%%%%%%%%%%%%%%%%%%%%%%%%%%%%%%%%%%%%%%%%%%%%%%%%%%%%%%%%%%%
%%%%%%%%%%%%%%%%%%%%%%%%%%%%%%%%%%%%%%%%%%%%%%%%%%%%%%%%%%%%%%%%%%%%%%%
\ack
The authors would like to thank the anonymous referee for useful
comments on the paper. EM acknowledge support for this study by the
Swedish Research Council and SIDA. CC is funded by the NRF (South
Africa).  This was work initiated at a meeting funded by the SASWE -
Cosmology Bilateral agreement between South Africa and Sweden.

%%%%%%%%%%%%%%%%%%%%%%%%%%%%%%%%%%%%%%%%%%%%%%%%%%%%%%%%%%%%%%%%%%%%%%%
%%%%%%%%%%%%%%%%%%%%%%%%%%%%%%%%%%%%%%%%%%%%%%%%%%%%%%%%%%%%%%%%%%%%%%%
%\section*{References}
\bibliographystyle{pippo1}
\bibliography{paper3}

\end{document}